\documentclass[twocolumn,english,english,nobibnotes,nofootinbib]{revtex4-1}
\usepackage[T1]{fontenc}
\usepackage[latin9]{inputenc}
\usepackage{float}
\usepackage{amsmath}
\usepackage{amssymb}
\usepackage{graphicx}

\makeatletter
\@ifundefined{textcolor}{}
{%
 \definecolor{BLACK}{gray}{0}
 \definecolor{WHITE}{gray}{1}
 \definecolor{RED}{rgb}{1,0,0}
 \definecolor{GREEN}{rgb}{0,1,0}
 \definecolor{BLUE}{rgb}{0,0,1}
 \definecolor{CYAN}{cmyk}{1,0,0,0}
 \definecolor{MAGENTA}{cmyk}{0,1,0,0}
 \definecolor{YELLOW}{cmyk}{0,0,1,0}
}

\usepackage[english]{babel}
\usepackage{braket}

\usepackage{tikz}\usepackage{mathtools}%

\usepackage{babel}
\usepackage{babel}

\usepackage{babel}

\makeatother

\usepackage{babel}
\begin{document}

\title{Theory of ballistic quantum transport in presence of localized defects}

\author{K. Kolasi\'{n}ski }

\affiliation{AGH University of Science and Technology, Faculty of Physics and
Applied Computer Science,\\
 al. Mickiewicza 30, 30-059 Kraków, Poland}

\author{A. Mre\'{n}ca-Kolasi\'{n}ska }

\affiliation{AGH University of Science and Technology, Faculty of Physics and
Applied Computer Science,\\
 al. Mickiewicza 30, 30-059 Kraków, Poland}

\author{B. Szafran}

\affiliation{AGH University of Science and Technology, Faculty of Physics and
Applied Computer Science,\\
 al. Mickiewicza 30, 30-059 Kraków, Poland}
\begin{abstract}
We present an efficient numerical approach for treating ballistic
quantum transport across devices described by tight binding (TB) Hamiltonians
designated to systems with localized potential defects. The method
is based on the wave function matching approach, Lippmann-Schwinger
equation (LEQ) and the scattering matrix formalism. We show that the
number of matrix elements of the Green's function to be evaluated
for the unperturbed system can be essentially reduced by projection
of the time reversed scattering wave functions on LEQ which radically
improves the speed and lowers the memory consumption of the calculations.
Our approach can be applied to quantum devices of an arbitrary geometry
and any number of degrees of freedom or leads attached. We provide
a couple of examples of possible applications of the theory, including
current equilibration at the p-n junction in graphene and scanning
gate microscopy mapping of electron trajectories in the magnetic focusing
experiment on a graphene ribbon. Additionally, we provide a simple
toy example of electron transport through 1D wire with added onsite
perturbation and obtain a simple formula for conductance showing that
Green's function of the device can be obtained from the conductance
versus impurity strength characteristics. 
\end{abstract}
\maketitle

\section{Introduction}

According to the Landauer approach the phase coherent component of
conductance in nanoscale and mesoscopic systems is determined by quantum
scattering of the electron incident from an input channel \cite{DattaBook}.
The coherent transport problem is of a nonlocal nature, as it is determined
by the electron wave function that is defined within with the entire
device with boundary conditions that are set at the ends of the sample.
Nevertheless, in a number of problems, the response of the wave function
to a local short-range perturbation is of a central interest for characterization
of the sample and its electrical properties. To name a few examples,
this is in particular the case for short range perturbations introduced
by the scanning techniques with a probe sweeping the surface of the
sample \cite{Sellier2011,Ferry2011,Topinka2000,Topinka2001,Kozikov2015,Jura2007,Jura2009},
for the scattering defects leading to the weak localization \cite{Altshuler1980}
and weak antilocalization effects \cite{Dresselhaus1992,Knap1996,Hassenkam1997},
or for lattice defects leading to valley mixing in graphene \cite{Morpurgo2006}.
Moreover, averaging over the coherent scatterers positions is one
of the numerical techniques to account for the decoherence effects
\cite{Pala2004}, equilibration of the currents in n-p-n junctions
in graphene \cite{Long2008} or investigation of Anderson localization
in graphene nanoribbons by introducing the disorder on ribbon edge
\cite{Libisch2012}.

Due to a nonlocality of quantum scattering the conductance response
of the system to a local perturbation calls for solution of the scattering
problem in the entire integration domain. For systems, in which the
perturbation can be separated from the Hamiltonian $\tilde{{\boldsymbol{H}}}={\boldsymbol{H}}+{\boldsymbol{V}}$,
one of the available procedures for finding the perturbed wave function
is a solution of the Lippman-Schwinger equation \cite{LippmannSchwinger}
spanned by the solution of the of the transport problem for $\boldsymbol{H}$,
the accompanied retarded Green's function and the perturbation operator
$\boldsymbol{V}$. In practice for electron transport problems the
Lippman-Schwinger equation is usually treated with the perturbation
expansion \cite{Jalabert2010,Gorini2013} or with iterative schemes
\cite{Sasakawa1983,Castelano2007}. In this paper we present a method
for an exact solution of the scattering problem with the Lippman-Schwinger
equation that requires evaluation of reduced Green's function matrix
elements defined within the region affected by the potential perturbation
only. The reduction is possible by projection of the Lippman-Schwinger
equation on the transport solutions with reversed time flow. The present
approach allows for a radical speed-up of the calculations whenever
various distributions of the perturbations $\boldsymbol{V}$ for the
same Hamiltonian $\boldsymbol{H}$ are needed. For illustration of
the method we present applications to scanning gate microscopy of
magnetic focusing \cite{Aidala2007} in graphene \cite{Bhandari2016,Taychatanapat2013},
and for evaluation of the fractional conductance plateaux for graphene
p-n junctions \cite{Barbaros2007,Long2008,Nakaharai2011} in the quantum
Hall regime \cite{Levitov2007,Williams2007}.

The paper is organized as follows. In the next Section we recall some
basics and introduce necessary quantities needed for quantum scattering
problem described within the robust and commonly used wave function
matching method \cite{Sorensen2009,Zwierzycki2008,Zwierzycki2006}.
In Section III we recall the tight binding version of Lippmann-Schwinger
equation. Next we show that by projecting the time reversed scattering
wave functions on the Lippmann-Schwinger equation we can significantly reduce the number of required
Green's functions elements, radically improving the memory consumption
and the speed of algorithm. Later we discuss a simple application
of our equation to the 1D model disordered by delta like impurity,
for which we show that the Green's function (the real part and the
imaginary part) can be computed from conductance impurity strength
characteristic. In Section IV we overview the established methods
for calculations of the Green's function of unperturbed systems, starting
from knitting algorithm for arbitrary shaped devices, fast recursive
equations for bulk materials or modular approach for creating structures.
In the last Section we show the examples of application of our method
for graphene based devices.

\section{The scattering approach for unperturbed Hamiltonian}

\begin{figure}[h]
\begin{centering}
\includegraphics[width=0.7\columnwidth]{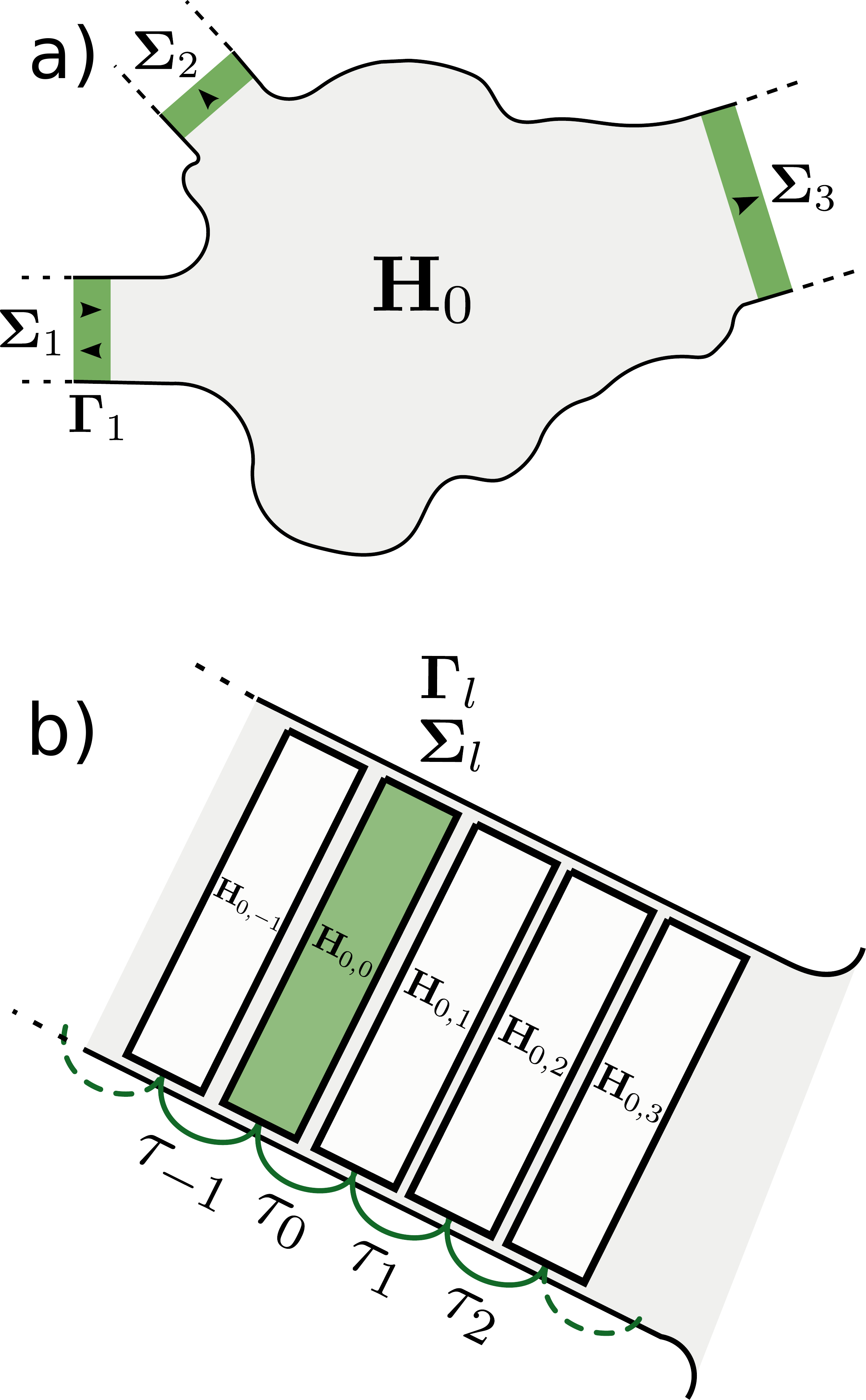} 
\par\end{centering}

\caption{\label{fig:sys}a) A schematic sketch of a quantum scatterer described by $\mathbf{H}_{0}$
coupled to the three semi-infinite leads by self-energy matrices $\mathbf{\Sigma}_{l}$.
The  electron comes from the source represented by $\Gamma_{l}$
source vector. The arrows point the possible direction of the scattering
electron for this specific example. b) Block tridiagonal partitioning
of the Hamiltonian near the leads from which $\mathbf{\Sigma}_{l}$
and $\mathbf{\Gamma}_{l}$ can be computed. The green area denotes the first
slice which belongs to the isolated system - the semi infinite lead
and quantum device interface atoms. }
\end{figure}

We start by describing the wavefunction matching approach \cite{Zwierzycki2008,Leng1994,Kirkner1990,Sorensen2009}
for solving scattering problem of arbitrary devices that we use to
solve the unperturbed problem for operator $H$ and that provides
the elements to define the method for treatment of the local perturbations.
We assume that the system of the interest can be expressed in terms
of finite size matrices like those generated by tight-binding (TB)
problems or finite difference approaches. The whole device can be
divided into two parts, the Hamiltonian of isolated system $\boldsymbol{H}_{0}$
and the self-energy $\boldsymbol{\Sigma}$ term which describes the
coupling of the isolated system to the semi-infinite channels (see
Fig. \ref{fig:sys}(a)). The total Hamiltonian is then defined as
\[
\boldsymbol{H}=\boldsymbol{H}_{0}+\boldsymbol{\Sigma},
\]
where the self-energy matrix contains the contribution from all the
leads connected to the device $\boldsymbol{\Sigma}=\boldsymbol{\Sigma}_{1}+\ldots+\boldsymbol{\Sigma}_{N}$.

In order to compute the self-energy matrix $\boldsymbol{\Sigma}_{l}$
for each lead $l$ one slices the Hamiltonian $\boldsymbol{H}_{0}$
at the lead interface into the block tridiagonal form (see Fig. \ref{fig:sys}(b))
\begin{equation}
-\boldsymbol{\tau}_{i}\boldsymbol{c}_{i-1}+\left(E_{\mathrm{F}}-\boldsymbol{H}_{0,i}\right)\boldsymbol{c}_{i}-\boldsymbol{\tau}_{i+1}^{\dagger}\boldsymbol{c}_{i+1}=\mathbf{0},\label{eq:Hc}
\end{equation}
where $i$ enumerates the $i-$th slice from the lead interface ($i=0$)
and $\boldsymbol{\tau}_{i}$ is the coupling matrix between two consecutive
Hamiltonian slices $\boldsymbol{H}_{0,i}$ and $\boldsymbol{H}_{0,i+1}$,
vector $\boldsymbol{c}_{i}$ is the wave function at slice $i$. Assuming
that the lead is homogeneous i.e. the Hamiltonian and the coupling
matrices do not depend on the position $i$ inside the lead, we may
drop the indices in Eq. (\ref{eq:Hc}) and obtain the formula 
\begin{equation}
-\boldsymbol{\tau}\boldsymbol{c}_{i-1}+\left(E_{\mathrm{F}}-\boldsymbol{H}_{0}\right)\boldsymbol{c}_{i}-\boldsymbol{\tau}^{\dagger}\boldsymbol{c}_{i+1}=\mathbf{0},\label{eq:Hclead}
\end{equation}
which can be solved numerically by applying Bloch substitution $\boldsymbol{c}_{i}=\lambda^{i}\boldsymbol{u}$
\cite{Zwierzycki2008,Rungger2008}, where $\lambda^{n}\equiv e^{ikn}$
describes the plane wave propagation in the channel with $k$ being
a wave vector and $\boldsymbol{u}$ -- is a Bloch mode which spans
the unit cell. The solution of Eq. (\ref{eq:Hclead}) leads to the
set of eigen pairs $\left(\left\{ \lambda_{1},\boldsymbol{u}_{1}\right\} ,\left\{ \lambda_{2},\boldsymbol{u}_{2}\right\} ,\ldots,\left\{ \lambda_{2N_{L}},\boldsymbol{u}_{2N_{L}}\right\} \right)$,
where $N_{L}$ is the length of the $\boldsymbol{c}_{i}$ vector.
Then we group $\left\{ \lambda_{i},\boldsymbol{u}_{i}\right\} $ into
$N_{L}$ incoming $\left\{ \lambda_{m,+},\boldsymbol{u}_{m,+}\right\} $
or $N_{L}$ outgoing $\left\{ \lambda_{m,-},\boldsymbol{u}_{m,-}\right\} $
modes. Each propagating mode $\boldsymbol{u}_{m,\pm}$ (i.e. with
$\left|\lambda_{m,\pm}\right|=1$) is then normalized to carry the
unit value of quantum flux \cite{Zwierzycki2008,Rungger2008}. We
define incoming/outgoing modes matrices as 
\[
\mathbf{U}_{\pm}=\left(\boldsymbol{u}_{1,\pm},\ldots,\boldsymbol{u}_{N_{L},\pm}\right)
\]
and diagonal $\mathbf{\Lambda}_{\pm}$ matrix constructed from Bloch
factors 
\[
\mathbf{\Lambda}_{\pm}=\left(\begin{array}{ccc}
\lambda_{1,\pm} &  & 0\\
 & \ddots\\
0 &  & \lambda_{N_{L},\pm}
\end{array}\right).
\]
Then the incoming/outgoing Bloch matrices are defined as 
\begin{equation}
\mathbf{F}_{\pm}\equiv\mathbf{U}_{\pm}\mathbf{\Lambda}_{\pm}^{-1}\mathbf{U}_{\pm}^{-1}.\label{eq:BlochF}
\end{equation}
Description of numerically stable algorithm for calculation of possibly
ill-conditioned $F_{\pm}$ can be found in Ref. \cite{Wimmer2009}.
Another approach which involves singular value decomposition (SVD)
method is explained in Ref. \cite{Rungger2008}. The self-energy matrix
$\mathbf{\Sigma}$ of a given lead $l$ is defined as 
\[
\mathbf{\Sigma}_{l}\equiv\mathbf{\tau}\mathbf{F}_{l,-}.
\]

A general expression for the scattering problem can be written in
terms of large but sparse system of linear equations 
\begin{equation}
\left(E_{\mathrm{F}}\boldsymbol{1}-\boldsymbol{H}\right)\boldsymbol{\Psi}_{l,m}=\boldsymbol{\Gamma}_{l,m},\label{eq:HYG}
\end{equation}
where $\mathbf{\Psi}_{l,m}$ is the scattering wavefunction of electron
incoming from lead $l$ in mode $m$ and $\boldsymbol{H}=\boldsymbol{H}_{0}+\boldsymbol{\Sigma}$.
The source vector $\boldsymbol{\Gamma}_{l,m}$ is non-zero only at
sites which belong to the lead $l$ and it is defined as 
\begin{equation}
\boldsymbol{\Gamma}_{l,m}=\boldsymbol{\tau}_{l}\left(\boldsymbol{F}_{l+}-\boldsymbol{F}_{l-}\right)\ket{\boldsymbol{u}_{l,m,+}}.\label{eq:SourceVec}
\end{equation}

After solution of the scattering problem (\ref{eq:HYG}) for a given
$m$-th incoming mode one may calculate transmission amplitudes from
\begin{equation}
\boldsymbol{t}_{l,m}^{l'}=\boldsymbol{U}_{l',-}^{-1}\boldsymbol{\Psi}_{l,m}^{L'},\label{eq:tm}
\end{equation}
and reflection amplitudes as 
\begin{equation}
\boldsymbol{r}_{l,m}^{l}=\boldsymbol{U}_{l,-}^{-1}\left(\boldsymbol{\Psi}_{l,m}^{L}-\boldsymbol{u}_{l,m,+}\right),\label{eq:rm}
\end{equation}
with $\boldsymbol{U}_{l,-}$ and $\boldsymbol{U}_{l',-}$ being the
outgoing modes matrices for input lead $l$ and output leads $l'$,
respectively \cite{Zwierzycki2008}. The superscripts $L',\, L$ written
by uppercase letters denote the set of elements of vector $\boldsymbol{\Psi}_{l,m}$
which belong to the leads $l'$ or $l$, respectively. We define $t_{l,m}^{l',m'}/r_{l,m}^{l',m'}$
as the transmission/reflection amplitude that electron entering the
device at lead $l$ in mode $m$ will leave the system at lead $l'$
in mode $m'$. The transmission/reflection vectors are denoted as
$\boldsymbol{t}_{l,m}^{\mathrm{l'}}$ and $\boldsymbol{r}_{l,m}^{\mathrm{l'}}$,
respectively. Having computed $\boldsymbol{t}$ and $\boldsymbol{r}$
amplitudes one calculates transport properties of the system: the
electrical conductance, the shot noise or thermoelectric coefficients.
For instance the differential conductance at $T=0$ can be computed
from the Landauer-Büttiker formula 
\begin{equation}
G_{l}^{l'}=\frac{e^{2}}{h}\sum_{m,m'}\left|t_{l,m}^{\mathrm{l',m'}}\right|^{2},\label{eq:G}
\end{equation}
where the sum over $m$ and $m'$ runs only through propagating modes
in lead $l$ and $l'$.

\section{The transport with localized disorder}

\subsection{The disorder matrix}

We are looking for a solution of the scattering problem of a system
distorted by some potential energy operator $\boldsymbol{V}$ 
\begin{equation}
\tilde{\boldsymbol{H}}=\boldsymbol{H}+\boldsymbol{V},\label{eq:Htilde}
\end{equation}
where we assume that $\boldsymbol{V}$ has following properties: 
\begin{enumerate}
\item $\boldsymbol{V}$ affects only a fraction of sites (and orbitals)
$P$ of whole Hamiltonian $\boldsymbol{H}$, i.e. $\left|V_{p,q}\right|\neq0$
for $p,q\in P$. The performance and memory consumption of the numerical
method derived below highly depends on the cardinal number of the
of $P$ set and will be discussed later. We denote by $\boldsymbol{V}_{PP}$
reduced $\boldsymbol{V}$ matrix of size $N_{P}=n(P)$ defined as
$\left[\boldsymbol{V}_{PP}\right]_{p,q}=V_{m(p),m(q)}$, where $\mathrm{m}(p)$
is a function which maps from the local indices of $\boldsymbol{V}_{PP}$
matrix i.e. from $p,q\in\{1,2,\ldots,N_{P}\}$ to global indices $m(p),m(q)\in P$
of larger $\boldsymbol{V}$ matrix. 
\item In general $\boldsymbol{V}_{PP}$ can be dense, complex but must be
hermitian ($\boldsymbol{V}=\boldsymbol{V}^{\dagger}$) in order to
conserve the current in the system. Note that the diagonal elements
of the potential matrix correspond to the on-site electrostatic potential
energy, where the off-diagonal elements correspond to hopping energies
between different sites. 
\item $\boldsymbol{V}$ does not affect the sites belonging to the leads
i.e. $\boldsymbol{V}$ does not change the modes in the leads. 
\end{enumerate}

\subsection{Lippmann Schwinger equation}

In this Section we specify the Lippmann-Schwinger equation for scattering
processes and discuss it possible application for TB systems \cite{Cerda1997STM}.
The perturbed scattering wavefunction $\tilde{\boldsymbol{\Psi}}_{l,m}$
for the system described with Eq. (\ref{eq:Htilde}) satisfies scattering
equation Eq. (\ref{eq:HYG})

\begin{equation}
\left(E_{\mathrm{F}}\boldsymbol{1}-\tilde{\boldsymbol{H}}\right)\tilde{\boldsymbol{\Psi}}_{l,m}=\boldsymbol{\Gamma}_{l,m}.\label{eq:HYG-1}
\end{equation}
Note that the source vector $\boldsymbol{\Gamma}_{l,m}$ does not
change, which results from the property (3) of the potential matrix
$\boldsymbol{V}$. Without loss of generality we express the new scattering
wavefunction in terms of the unperturbed one 
\[
\tilde{\boldsymbol{\Psi}}_{l,m}=\boldsymbol{\Psi}_{l,m}+\delta\boldsymbol{\Psi}_{l,m}
\]
then using the Eq. (\ref{eq:HYG}) and (\ref{eq:HYG-1}) we get

{\footnotesize{{ 
\begin{eqnarray}
\left(E_{\mathrm{F}}\boldsymbol{1}-\left(\boldsymbol{H}+\boldsymbol{V}\right)\right)\left(\boldsymbol{\Psi}_{l,m}+\delta\boldsymbol{\Psi}_{l,m}\right) & = & \boldsymbol{\Gamma}_{l,m},\nonumber \\
\left(E_{\mathrm{F}}\boldsymbol{1}-\boldsymbol{H}\right)\delta\boldsymbol{\Psi}_{l,m} & = & \boldsymbol{V}\left(\boldsymbol{\Psi}_{l,m}+\delta\boldsymbol{\Psi}_{l,m}\right).\label{eq:p1}
\end{eqnarray}
}}}The $\boldsymbol{G}=\left(E_{\mathrm{F}}\boldsymbol{1}-\boldsymbol{H}\right)^{-1}$
is the Green's function of the unperturbed system. By multiplying
Eq. (\ref{eq:p1}) from the left by $\boldsymbol{G}$ we obtain the
Lippmann-Schwinger equation for $\delta\boldsymbol{\Psi}_{\mathrm{l,m}}$
\begin{equation}
\delta\boldsymbol{\Psi}_{l,m}=\boldsymbol{G}\boldsymbol{V}\left(\boldsymbol{\Psi}_{l,m}+\delta\boldsymbol{\Psi}_{l,m}\right).\label{eq:dY}
\end{equation}
In order to compute the conductance $G_{l\rightarrow l'}$ (see Eq.
(\ref{eq:tm}) and (\ref{eq:G})) of the system with $\boldsymbol{V}$
one has to evaluate the values of $\tilde{\boldsymbol{\Psi}}_{\mathrm{l,m}}$
at each lead $l'$ interface i.e. vectors $\tilde{\boldsymbol{\Psi}}_{l,m}^{L'}$,
thus we only need to compute the elements of the $\delta\Psi_{l,m}$
vector which belong to the given lead. The change in the wave function
at site $l'$ is given by 
\begin{eqnarray*}
\delta\Psi_{l,m}^{l'} & = & \sum_{p,q}G_{l',p}V_{p,q}\left(\Psi_{l,m}^{q}+\delta\Psi_{l,m}^{q}\right).
\end{eqnarray*}
Using the first property of the $V$ matrix we get

\begin{eqnarray*}
\delta\Psi_{l,m}^{l'} & = & \sum_{p,q\in P}G_{l',p}V_{p,q}\left(\Psi_{l,m}^{q}+\delta\Psi_{l,m}^{q}\right)
\end{eqnarray*}
which in the matrix notation can be written as

\begin{eqnarray}
\delta\boldsymbol{\Psi}_{l,m}^{L'} & = & \boldsymbol{G}_{L'P}\boldsymbol{V}_{PP}\left(\boldsymbol{\Psi}_{l,m}^{P}+\delta\boldsymbol{\Psi}_{l,m}^{P}\right),\label{eq:dYL}
\end{eqnarray}
where $\boldsymbol{G}_{L'P}$ is the reduced Green's matrix which
couples the elements of the lead $l'$ with the perturbed sites $P$.
Similarly, using Eq. (\ref{eq:dY}) we calculate $\delta\boldsymbol{\Psi}_{l,m}^{P}$
as 
\begin{eqnarray*}
\delta\boldsymbol{\Psi}_{l,m}^{P} & = & \boldsymbol{G}_{PP}\boldsymbol{V}_{PP}\left(\boldsymbol{\Psi}_{l,m}^{P}+\delta\boldsymbol{\Psi}_{l,m}^{P}\right)\\
 & = & \left(1-\boldsymbol{G}_{PP}\boldsymbol{V}_{PP}\right)^{-1}\boldsymbol{G}_{PP}\boldsymbol{V}_{PP}\boldsymbol{\Psi}_{l,m}^{P},
\end{eqnarray*}
which substituted to Eq. (\ref{eq:dYL}) gives the final formula for
the change in the wave function{\footnotesize{{ 
\begin{eqnarray}
\delta\boldsymbol{\Psi}_{l,m}^{L'} & = & \boldsymbol{G}_{L'P}\boldsymbol{V}_{PP}\left(\boldsymbol{1}+\left(1-\boldsymbol{G}_{PP}\boldsymbol{V}_{PP}\right)^{-1}\boldsymbol{G}_{PP}\boldsymbol{V}_{PP}\right)\boldsymbol{\Psi}_{l,m}^{P},\nonumber \\
 & = & \boldsymbol{G}_{L'P}\boldsymbol{V}_{PP}\left(1-\boldsymbol{G}_{PP}\boldsymbol{V}_{PP}\right)^{-1}\boldsymbol{\Psi}_{l,m}^{P}\nonumber \\
 & \equiv & \boldsymbol{G}_{L'P}\boldsymbol{T}_{PP}\boldsymbol{\Psi}_{l,m}^{P},\label{eq:dYf}
\end{eqnarray}
}}}where 
\begin{equation}
\boldsymbol{T}_{PP}=\boldsymbol{V}_{PP}\left(1-\boldsymbol{G}_{PP}\boldsymbol{V}_{PP}\right)^{-1}\label{eq:Tmat}
\end{equation}
is the transition matrix. Then the transmission probability through
lead $l'$ can be computed from Eq. (\ref{eq:tm}) 
\[
\tilde{\boldsymbol{t}}_{l,m}^{l'}=\boldsymbol{U}_{l',-}^{-1}\left(\boldsymbol{\Psi}_{l,m}^{L'}+\delta\boldsymbol{\Psi}_{l,m}^{L'}\right).
\]
Before we proceed to further simplification of Eq. (\ref{eq:dYf}),
let us discuss some of numerical properties of the obtained result.
Firstly, in order to find the transmission probabilities one has to
compute the selected elements of the Green's function of unperturbed
system $\boldsymbol{G}$, which are needed to construct two reduced
matrices: $\boldsymbol{G}_{L'P}$ and $\boldsymbol{G}_{PP}$. We will
discuss this problem in the next Sections but for now we assume that
those matrices can be computed with available algorithms. Secondly,
having $\boldsymbol{G}_{L'P}$ and $\boldsymbol{G}_{PP}$ one may
compute conductance without solving large system of linear equations
(\ref{eq:HYG}) which allows for speed up of the calculations. The
speed of the algorithm will depend on the time $T_{P}$ needed to
calculate the $\boldsymbol{G}_{L'P}\boldsymbol{T}_{PP}$ matrix, time
$T_{G}$ needed for calculation of the reduced Green's functions and
time $T_{W}$ needed for calculation of the unperturbed wave functions.
On the other hand the computation time for the standard method is
only $T_{W}$ since the perturbation matrix $\mathbf{V}$ does not
change the solution time of Eq. \eqref{eq:HYG-1}. Hence, for a single
scattering process the proposed method is slower by ratio $(T_{P}+T_{G}+T_{W})/T_{W}$.
However, when one is interested in statistical properties of current
and needs to compute conductance $N$ times for different $\boldsymbol{V}_{PP}$
matrices (here we assume that the set of perturbed sites $P$ does
not change i.e. we can compute $\boldsymbol{G}_{L'P}$ and $\boldsymbol{G}_{PP}$
once and store them in memory) the ratio becomes $(NT_{P}+T_{G}+T_{W})/\left(NT_{W}\right)$
and for $N\rightarrow\infty$ leads to $T_{P}/T_{W}$, which in general
can be an arbitrarily small number. Note that the time $T_{P}$ depends
only on size of the $\boldsymbol{V}_{PP}$ matrix which means that
for small $V_{PP}$ i.e. $\leq1000$ finding the transmission amplitudes
for a system with $10^{6}$ sites may be significantly reduced from
minutes to fraction of seconds. Finally, the form of Eq. (\ref{eq:dYf})
requires allocation of several dense matrices, two of size $N_{P}\times N_{P}$
($\boldsymbol{G}_{PP}$ and $\boldsymbol{V}_{PP}$) and $L$ matrices
of size $N_{L'}\times N_{P}$ ($\boldsymbol{G}_{L'P}$), where $N_{L'}$
is the total number of elements in the lead $l'$ vector and $L$
is the total number of leads in the system. The value of $N_{P}$
can be controlled by choosing the number of disordered sites in the
system, however the $N_{L'}$ depends on the device structure and
can be in general very large increasing the memory usage and the times
$T_{G}$, $T_{P}$. We propose then to use Eq. (\ref{eq:dYf}) as
a starting point for more complicated approach discussed in the next
paragraph.

\subsection{Excluding the $\boldsymbol{G}_{L'P}$ terms}

In this Section we show that the $\boldsymbol{G}_{L'P}$ term can
be eliminated from Eq. (\ref{eq:dYf}). Firstly, let us note that
another matrix $\boldsymbol{G}_{PL'}$ can be related with the scattering
wave function at the sites $P$ with equation 
\[
\boldsymbol{\Psi}_{l,m}^{P}=\boldsymbol{G}_{PL}\boldsymbol{\Gamma}_{l,m}^{L},
\]
where we used Eq. (\ref{eq:HYG}) and the sparsity  of the
source vector i.e. $\boldsymbol{\Gamma}_{l,m}^{Q}=0$ for $Q\notin L$.
Now, we look for similar relation for $\boldsymbol{G}_{L'P}$. Firstly,
we note that by conjugating Hamiltonian $\boldsymbol{H}_{0}\rightarrow\boldsymbol{H}_{0}^{*}=\boldsymbol{H}_{0}^{T}$
in Eq. (\ref{eq:HYG}) we obtain scattering problem for the particle
propagating backward in time 
\begin{equation}
\boldsymbol{\overleftarrow{\Psi}}_{l',n}=\overleftarrow{\boldsymbol{G}}\boldsymbol{\overleftarrow{\Gamma}}_{l',n},\label{eq:GammaL}
\end{equation}
where the symbol $\overleftarrow{X}\equiv X\left[\boldsymbol{H}_{0}^{*}\right]$
denotes that variable $X$ is computed as usual but for conjugated
Hamiltonian matrix $\boldsymbol{H}_{0}^{*}$, the $\boldsymbol{\overleftarrow{\Psi}}_{l',n}$
and $\boldsymbol{\overleftarrow{\Gamma}}_{l',n}$ are the column vectors
and the Green's function $\overleftarrow{\boldsymbol{G}}$ is calculated
as 
\[
\overleftarrow{\boldsymbol{G}}=\frac{1}{E_{\mathrm{F}}\boldsymbol{1}-\left(\boldsymbol{H}_{0}^{*}+\overleftarrow{\boldsymbol{\Sigma}}\right)}.
\]
Additionally, since $\boldsymbol{H}_{0}^{*}=\boldsymbol{H}_{0}^{T}$
one may prove that $\overleftarrow{\boldsymbol{\Sigma}}=\boldsymbol{\Sigma}^{T}$
which leads to following identity $\overleftarrow{\boldsymbol{G}}^{T}=\boldsymbol{G}$.
Using this relation and transposing Eq. (\ref{eq:GammaL}) we may
write 
\[
\boldsymbol{\overleftarrow{\Psi}}_{l',n}^{(T)}=\boldsymbol{\overleftarrow{\Gamma}}_{l',n}^{(T)}\overleftarrow{\boldsymbol{G}}^{T}=\boldsymbol{\overleftarrow{\Gamma}}_{l',n}^{(T)}\boldsymbol{G},
\]
where for clarity we denote transposition as $(T)$ in order to distinguish
it from other superscripts. Evaluating this expression at sites belonging
to $P$ set we get 
\begin{equation}
\boldsymbol{\overleftarrow{\Psi}}_{l',n}^{P(T)}=\boldsymbol{\overleftarrow{\Gamma}}_{l',n}^{L'(T)}\boldsymbol{G}_{L'P}.\label{eq:Yback}
\end{equation}
Let us now project the $\boldsymbol{\overleftarrow{\Gamma}}_{l',n}^{L'(T)}$
vectors on the scattering wavefunction at lead $l'$ (see Eq. \ref{eq:dYf})
\[
\tilde{\boldsymbol{\Psi}}_{l,m}^{L'}-\boldsymbol{\Psi}_{l,m}^{L'}=\boldsymbol{G}_{L'P}\boldsymbol{T}_{PP}\boldsymbol{\Psi}_{l,m}^{P}.
\]
Using Eq. \eqref{eq:Yback} we get

\begin{eqnarray*}
\boldsymbol{\overleftarrow{\Gamma}}_{l',n}^{L'(T)}\left(\tilde{\boldsymbol{\Psi}}_{l,m}^{L'}-\boldsymbol{\Psi}_{l,m}^{L'}\right) & = & \boldsymbol{\overleftarrow{\Psi}}_{l',n}^{P(T)}\boldsymbol{T}_{PP}\boldsymbol{\Psi}_{l,m}^{P}.
\end{eqnarray*}
The equation above can be related with transmission probabilities
(\ref{eq:tm}) 
\begin{equation}
\boldsymbol{\overleftarrow{\Gamma}}_{l',n}^{L'(T)}\boldsymbol{U}_{l',-}\delta\boldsymbol{t}_{l,m}^{l'}=S_{l,m}^{l',n},\label{eq:GUt}
\end{equation}
where we define the elements of the scattering overlap matrix $\boldsymbol{S}_{l,m}^{l'}$
\[
S_{l,m}^{l',n}\equiv\boldsymbol{\overleftarrow{\Psi}}_{l',n}^{P(T)}\boldsymbol{T}_{PP}\boldsymbol{\Psi}_{l,m}^{P},
\]
and the variation in the transmission vector 
\begin{eqnarray}
\delta\boldsymbol{t}_{l,m}^{l'} & \equiv\tilde{\boldsymbol{t}}_{l,m}^{l'}-\boldsymbol{t}_{l,m}^{l'}= & \boldsymbol{U}_{l',-}^{-1}\left(\tilde{\boldsymbol{\Psi}}_{l,m}^{L'}-\boldsymbol{\Psi}_{l,m}^{L'}\right)\label{eq:dt}
\end{eqnarray}
Let us now discuss the dimensions of the vectors and matrices present
in the equations above. The $\boldsymbol{\Psi}_{l,m}^{P}$ is a column
vector of length $N_{P}$. We define a matrix 
\begin{equation}
\boldsymbol{\Psi}_{M_{L}}^{P}\equiv\left(\boldsymbol{\Psi}_{l,1}^{P},\boldsymbol{\Psi}_{l,2}^{P},\ldots,\boldsymbol{\Psi}_{l,M_{L}}^{P}\right),\label{eq:YPML}
\end{equation}
composed from vectors $\boldsymbol{\Psi}_{l,m}^{P}$, where the uppercase
subscript $M_{L}$ denotes the number of propagating modes in lead
$l$ at given $E_{\mathrm{F}}$. Hence $\boldsymbol{\Psi}_{M_{L}}^{P}$
is a rectangular matrix of size $N_{P}\times M_{L}$. Similarly, for
the rest of the leads $l'$, we define 
\begin{equation}
\boldsymbol{\overleftarrow{\Psi}}_{M_{L'}}^{P(T)}\equiv\left(\overleftarrow{\boldsymbol{\Psi}}_{l',1}^{P},\overleftarrow{\boldsymbol{\Psi}}_{l',2}^{P},\ldots,\overleftarrow{\boldsymbol{\Psi}}_{l',M_{L'}}^{P}\right)^{T}\label{eq:YPMLprim}
\end{equation}
being a matrix of size $M_{L'}\times N_{P}$. Let us now define a
source matrix for modes propagating from lead $l'$ backward in time
\begin{equation}
\boldsymbol{\overleftarrow{\Gamma}}_{M_{L'}}^{L'(T)}\equiv\left(\boldsymbol{\overleftarrow{\Gamma}}_{l',1}^{L'},\boldsymbol{\overleftarrow{\Gamma}}_{l',2}^{L'},\ldots,\boldsymbol{\overleftarrow{\Gamma}}_{l',M_{L'}}^{L'}\right)^{T}\label{eq:GammaMLprim}
\end{equation}
The size of this matrix is $M_{L'}\times N_{L'}$. Finally, the modes
matrices $\boldsymbol{U}_{l',-}$ and the transmission vectors $\tilde{\boldsymbol{t}}_{l,m}^{l'}$
have dimensions $N_{L'}\times N_{L'}$ and $N_{L'}\times1$, respectively.

Let us note that $\boldsymbol{t}_{l,m}^{l'}$ is a vector which in
general can be divided into two parts i) scattering amplitudes of
propagating modes ii) and evanescent modes. Despite the fact that
the second term does not contribute to the current in the Landauer
formula the coefficients are usually non zero and play an important role
for construction of the open boundary conditions at leads interfaces.
The structure of the $\boldsymbol{t}$ vector can be written in general
form 
\begin{eqnarray*}
\boldsymbol{t}_{l,m}^{l'} & = & \left(t_{l,m}^{l',1},\ldots,t_{l,m}^{l',M_{L'}},t_{l,m}^{l',M_{L'}+1},\ldots,t_{l,m}^{l',N_{L'}}\right)^{T}.
\end{eqnarray*}
In the following we assume that the disorder introduced by $\boldsymbol{V}_{PP}$
matrix does not affect the evanescent modes at each lead or it may
be neglected, which allows us to write $\tilde{\boldsymbol{t}}_{l,m}^{l'}$
as 
\begin{equation}
\tilde{\boldsymbol{t}}_{l,m}^{l'}\equiv\left(\tilde{t}_{l,m}^{l',1},\ldots,\tilde{t}_{l,m}^{l',M_{L'}},t_{l,m}^{l',M_{L'}+1},\ldots,t_{l,m}^{l',N_{L'}}\right)^{T},\label{eq:assump}
\end{equation}
where the evanescent part of the $\tilde{\boldsymbol{t}}_{l,m}^{l'}$
is the same as in $\boldsymbol{t}_{l,m}^{l'}$ vector. From the above
we get the change in the scattering amplitudes 
\begin{eqnarray*}
\delta\boldsymbol{t}_{l,m}^{l'} & = & \left(\delta t_{l,m}^{l',1},\ldots,\delta t_{l,m}^{l',M_{L'}},0,\ldots,0\right)^{T},\\
 & \equiv & \left(\delta\tilde{\boldsymbol{t}}_{l,m}^{l'},\boldsymbol{0}_{1\times N_{L'}-M_{L'}}\right)^{T}
\end{eqnarray*}
which shows that we can truncate last $N_{L'}-M_{L'}$ columns of
modes matrix $\boldsymbol{U}_{\mathrm{l'},-}$ in Eq. (\ref{eq:GUt}).
Let us then denote the truncated matrix $\boldsymbol{U}_{\mathrm{l'},-}$
as $\boldsymbol{U}_{l',-}^{\mathrm{trnc}}$ which has dimensions $N_{L'}\times M_{L'}$
and is obtained from the first $M_{L'}$ columns of $\boldsymbol{U}_{l',-}$.
Using the definitions above we can write Eq. (\ref{eq:dt}) in the
following way 
\[
\boldsymbol{\overleftarrow{\Gamma}}_{M_{L'}}^{L'(T)}\boldsymbol{U}_{l',-}^{\mathrm{trnc}}\delta\tilde{\boldsymbol{t}}_{l,m}^{l'}=\boldsymbol{S}_{l,m}^{l'},
\]
where the product 
\begin{equation}
\overleftarrow{\boldsymbol{D}}_{l'}\equiv\boldsymbol{\overleftarrow{\Gamma}}_{M_{L'}}^{L'(T)}\boldsymbol{U}_{l',-}^{\mathrm{trnc}}\label{eq:Dlprim}
\end{equation}
is now a square matrix of size $M_{L'}\times M_{L'}$. Multiplying
the equation above by $\overleftarrow{\boldsymbol{D}}_{l'}^{-1}$
from the left we get 
\[
\delta\tilde{\boldsymbol{t}}_{l,m}^{l'}=\overleftarrow{\boldsymbol{D}}_{l'}^{-1}\boldsymbol{S}_{l,m}^{l'},
\]
from which one may compute transmission probabilities for propagating
modes 
\begin{equation}
\tilde{\boldsymbol{T}}_{l,m}^{l'}\equiv\left|\tilde{\boldsymbol{t}}_{l,m}^{l'}\right|^{2}=\left|\boldsymbol{t}_{l,m}^{l'}+\overleftarrow{\boldsymbol{D}}_{l'}^{-1}\boldsymbol{S}_{l,m}^{l'}\right|^{2},\label{eq:Ttilde}
\end{equation}
and conductance $\tilde{G}_{l}^{l'}$ (\ref{eq:G}). When $l=l'$
one obtains the reflection probabilities.

\subsection{Numerical algorithm}

To conclude the previous Section the following algorithm can be used
to compute the scattering matrix for system with potential matrix
$\boldsymbol{V}$: 
\begin{enumerate}
\item Compute and store the following matrices for system without disorder
potential (see Eq. (\ref{eq:HYG})): a) the scattering wave functions
$\boldsymbol{\Psi}_{l,m}$ for selected leads and modes in those leads.
b) the outgoing modes matrices $\boldsymbol{U}_{l',-}$ and $\boldsymbol{r}_{l,m}^{l}$(i.e.
scattering matrix). 
\item Compute and store the following matrices for system with $\boldsymbol{H}_{0}\rightarrow\boldsymbol{H}_{0}^{*}$
in Eq. (\ref{eq:HYG}) a) all the scattering wave functions $\overleftarrow{\boldsymbol{\Psi}}_{l',n}$
b) and source vectors $\boldsymbol{\overleftarrow{\Gamma}}_{l',n}^{L'}$
for time reversed problem. 
\item Compute reduced matrices $\boldsymbol{\Psi}_{M_{L}}^{P}$ from Eq.
(\ref{eq:YPML}), $\boldsymbol{\overleftarrow{\Psi}}_{M_{L'}}^{P(T)}$
from Eq. (\ref{eq:YPMLprim}), $\boldsymbol{\overleftarrow{\Gamma}}_{M_{L'}}^{L'(T)}$
from Eq. (\ref{eq:GammaMLprim}) and $\overleftarrow{\boldsymbol{D}}_{l'}$
from Eq. (\ref{eq:Dlprim}). 
\item Calculate the selected elements of the Green's function $\boldsymbol{G}_{PP}$
and reduced disorder matrix $\boldsymbol{V}_{PP}$. 
\item Construct the transition matrix $\boldsymbol{T}_{PP}$ from Eq. (\ref{eq:Tmat}). 
\item Compute new scattering amplitudes $\tilde{\boldsymbol{t}}_{l,m}^{l'}$
from Eq. (\ref{eq:Ttilde}) from which conductance (\ref{eq:G}) can
be calculated. 
\end{enumerate}

\subsection{Discussion}

The general relation between the scattering matrix and the transition
matrix (\ref{eq:Ttilde}) can be used for any type of scattering problem
which can be described by Eq. (\ref{eq:HYG}), hence it is relevant
for any kind of TB systems or e.g. Hamiltonians generated by finite
difference (or finite elements \cite{Leng1994,Kramer2013}) methods.
Then different kind of forms of $\boldsymbol{V}_{PP}$ can be used
to simulate random onsite potential in quantum structures (diagonal
form of $\boldsymbol{V}_{PP}$), point defects in the lattice, adatoms
\cite{CastroNeto2009,Fabian2015,Fabian2013} or modification of existing
hoping energies. The advantage of the main Eq. (\ref{eq:Ttilde})
over the basic Lippmann-Schwinger Eq. (\ref{eq:dYf}) is that one
reduces the number of Green's function elements to be computed, which
for arbitrary systems can be memory and time consuming task. On the
other hand one has to compute all the scattering wave functions for
particle propagating backward in time $\overleftarrow{\boldsymbol{\Psi}}_{l',n}$,
however this can be done efficiently with existing numerical libraries
\cite{superlu99,PARDISO}. Additionally, for well written quantum
transport solvers this problem reduce to replacing the original Hamiltonian
by its conjugation which is a straight forward task. One should also
note that in a special case when $\boldsymbol{H}_{0}$ is real the
relation $\overleftarrow{\boldsymbol{X}}=\boldsymbol{X}$ holds, hence
one does not have to compute $\overleftarrow{\boldsymbol{\Psi}}_{l',n}$
and other matrices separately.

\subsection{Weak perturbation limit}

An interesting case arises when one takes the limit $\boldsymbol{G}_{PP}\boldsymbol{V}_{PP}\rightarrow0$
in Eq. (\ref{eq:Tmat}) i.e. $\boldsymbol{V}_{PP}$ generates weak
perturbation in the Hamiltonian 
\begin{eqnarray*}
\boldsymbol{T}_{PP} & = & \boldsymbol{V}_{PP}\left(1-\boldsymbol{G}_{PP}\boldsymbol{V}_{PP}\right)^{-1}\approx\boldsymbol{V}_{PP}\left(1-\boldsymbol{G}_{PP}\boldsymbol{V}_{PP}\right)\\
 & = & \boldsymbol{V}_{PP}-\boldsymbol{V}_{PP}\boldsymbol{G}_{PP}\boldsymbol{V}_{PP},
\end{eqnarray*}
In this limit one may compute the correction to the scattering matrix
as 
\begin{eqnarray*}
\tilde{\boldsymbol{t}}_{l,m}^{l'} & = & \boldsymbol{t}_{l,m}^{l'}+\overleftarrow{\boldsymbol{D}}_{l'}^{-1}\boldsymbol{\overleftarrow{\Psi}}_{M_{L'}}^{P(T)}\boldsymbol{T}_{PP}\boldsymbol{\Psi}_{M_{L}}^{P}\\
 & \approx & \boldsymbol{t}_{l,m}^{l'}+\boldsymbol{t}_{l,m}^{l'(1)}+\boldsymbol{t}_{l,m}^{l'(2)},
\end{eqnarray*}
with the first (1) and the second (2) order corrections being 
\begin{eqnarray}
\boldsymbol{t}_{l,m}^{l'(1)} & = & \overleftarrow{\boldsymbol{D}}_{l'}^{-1}\boldsymbol{\overleftarrow{\Psi}}_{M_{L'}}^{P(T)}\boldsymbol{V}_{PP}\boldsymbol{\Psi}_{M_{L}}^{P}\label{eq:tp1}\\
\boldsymbol{t}_{l,m}^{l'(2)} & = & -\overleftarrow{\boldsymbol{D}}_{l'}^{-1}\boldsymbol{\overleftarrow{\Psi}}_{M_{L'}}^{P(T)}\boldsymbol{V}_{PP}\boldsymbol{G}_{PP}\boldsymbol{V}_{PP}\boldsymbol{\Psi}_{M_{L}}^{P}.\nonumber 
\end{eqnarray}
Note that the first correction does not require the information about
Green's function. Similar expression for the first order correction
(\ref{eq:tp1}) to the scattering matrix has been derived recently
\cite{Jalabert2010} in the context of scanning gate microscopy technique.
The present result generalizes the ones of Ref. \cite{Jalabert2010}
to the case of magnetic field, the spin degree of freedom or any other
system that can be described within the single-electron transport
problem defined by Eq. \eqref{eq:HYG}.% \textcolor{red}{Nie wiem czy robi\'{c}
%dyskusje o LDOS i tak ju\.{z} jest du\.{z}o}

\subsection{1D case }

Let us discuss another interesting scenario of one dimensional quantum
transport with a single scattering mode in the leads. We show that
for the case when the perturbation potential is localized on one site
$P$ ($\delta$ like potential) we can find simple expression for
the conductance which relate $G$ with Green's function at site $P$.
We can drop all the indices in Eq. (\ref{eq:Ttilde}) and work only
with scalar variables

\begin{eqnarray}
\tilde{t} & = & t+c_{P}V_{P}\left(1-G_{PP}V_{P}\right)^{-1}\label{eq:t1d}
\end{eqnarray}
with $c_{P}=\overleftarrow{D}^{-1}\overleftarrow{\Psi}^{P}\Psi^{P}$
and $V_{P}$ being the onsite potential affecting site $P$. For an
infinite potential barrier $V_{P}=+\infty$ the scattered electron
will be completely reflected, hence 
\[
\lim_{V_{P}\rightarrow+\infty}\tilde{t}=0=t-c_{P}G_{PP}^{-1}\Rightarrow t=c_{P}G_{PP}^{-1}.
\]
Using this result we simplify Eq. (\ref{eq:t1d}) 
\begin{eqnarray*}
\tilde{t} & = & c_{P}G_{PP}^{-1}+\frac{c_{P}V_{P}}{1-G_{PP}V_{P}}\\
 & = & \frac{c_{P}G_{PP}^{-1}}{1-G_{PP}V_{P}}=\frac{t}{1-G_{PP}V_{P}},
\end{eqnarray*}
from which we can compute the two terminal conductance 
\begin{equation}
\tilde{G}=G\frac{1}{1+\left[\Re\left\{ G_{PP}\right\} ^{2}+\Im\left\{ G_{PP}\right\} ^{2}\right]V_{P}^{2}-2V_{P}\Re\left\{ G_{PP}\right\} }.\label{eq:G1dresponse}
\end{equation}
From the equation above we see that by adding a localized potential
at some point $P$ to the quantum wire we can perform scan in function
of the $V_{P}$ amplitude and then fit the obtained response $\tilde{G}$
to the Eq. (\ref{eq:G1dresponse}) in order to obtain the information
about the Green's function (real and imaginary part i.e. LDOS) in
the device at point $P$. Another approach would be to calculate numerically
the first and second derivative of $\tilde{G}$ with respect to perturbation
strength $V_{P}$ 
\begin{eqnarray*}
\left.\frac{1}{G}\frac{d\tilde{G}}{dV_{P}}\right|_{V_{P}=0} & = & 2\Re\left\{ G_{PP}\right\} \\
\left.\frac{1}{G}\frac{d^{2}\tilde{G}}{dV_{P}^{2}}\right|_{V_{P}=0} & = & 6\Re\left\{ G_{PP}\right\} ^{2}-2\Im\left\{ G_{PP}\right\} ^{2}.
\end{eqnarray*}

Expression (\ref{eq:G1dresponse}) is exact for delta like perturbations,
however it should be also valid for finite size potentials when the
effective width of perturbation is smaller than half of the Fermi
wavelength $d_{V}\leq\lambda_{F}/2$. %Otherwise one could try to derive
%more general expression which would include this effect.

\section{Computing the Green's function}

In this Section we overview the procedures used for evaluation of
the Green's function matrices, for a general case (IV.A) and for a
devices with translational symmetry (IV.B) and the combined modular
approach for calculation of the Green's function (IV.C).

\subsection{Computing selected elements of $\boldsymbol{G}$ for arbitrary devices}

One of the most challenging aspects of the derived method is the calculation
of selected elements of the Green's function of unperturbed system
$\boldsymbol{G}=\left(E_{\mathrm{F}}\boldsymbol{1}-\boldsymbol{H}\right)^{-1}$.
Since $\boldsymbol{H}$ in usual applications is a large (e.g. $\sim10^{5}-10^{6}$)
and sparse matrix, its inverse is dense and cannot be computed with
direct inversion algorithms. In order to overcome this problem various
algorithms were developed to compute only the necessary elements of
$\boldsymbol{G}$ instead of whole matrix. A popular method called
recursive Green's function (RGF) which involves the Dyson equation
developed in a number of variants is used for this purpose e.g. see
Refs. \cite{Metalidis2005,Sanvito1999,Rotter2003,Guan2003}. Unfortunately
many of those variants are limited to a specific geometry of device
\cite{Settnes2015} or can be used only for two terminal devices or
special slicing has to be performed in order to include the effect
of multiterminal devices \cite{Thorgilsson2014256}.

However, recently a variation of RGF method has been developed which
generalizes the standard approach, leading to numerically stable knitting
algorithm \cite{Kazymyrenko2008}. The knitting\emph{ }algorithm can
be applied to arbitrary shaped devices with arbitrary number of leads,
orbitals etc. In general the method of Ref. \cite{Kazymyrenko2008}
can be used to compute selected elements of inverse of any structurally
symmetric sparse matrix. For more details about implementation, numerical
scaling or memory usage we refer the Reader to the original paper
\cite{Kazymyrenko2008}. Alternatively one may use efficient nested
dissection approach as described recently in Ref. \cite{Hetmaniuk2013}.

To summarize this Section, the selected elements of the Green's function
$\boldsymbol{G}$ can be computed with available algorithms, but for
general purpose it may be practical to implement an universal algorithm
from Ref. \cite{Kazymyrenko2008}. For testing purpose we provide
the source code of our implementation of the knitting algorithm \cite{Knitinv2016}
written in Fortran.

\subsection{Computing $\boldsymbol{G}$ of translational invariant devices}

In this Section we explain that for a special case of quantum channels
with translational symmetry (bulk materials) the general formula for
a Green's function can be found for any type of Hamiltonian including
the topology, number of orbitals or dimensionality of the problem
\cite{Sanvito1999,Zwierzycki2006,Libisch2012}.

Let us recall that any translationally invariant quantum channel generated
by the TB problems can be described by general block tridiagonal Hamiltonian
$E_{\mathrm{F}}\boldsymbol{1}-\boldsymbol{H}_{0}$ \cite{Khomaykov2005,Zwierzycki2008}
of form 
\[
-\boldsymbol{H}_{0i,i-1}\boldsymbol{c}_{i-1}+\left(E_{\mathrm{F}}\boldsymbol{1}-\boldsymbol{H}_{0i,i}\right)\boldsymbol{c}_{i}-\boldsymbol{H}_{0i,i+1}\boldsymbol{c}_{i+1}=\boldsymbol{0},
\]
where $\boldsymbol{c}_{i}$describe the wave function at $i$-th slice,
the diagonal block $\boldsymbol{H}_{0,i,i}$ is the Hamiltonian of
isolated slice and $\boldsymbol{H}_{0i,i-1}=\boldsymbol{H}_{0i,i+1}^{\dagger}=\boldsymbol{\tau}$
represent coupling between slices $i\mp1$ and $i$. Assumed translation
symmetry requires invariance of block matrices after shift along the
diagonal of $\boldsymbol{H}$ matrix 
\[
\boldsymbol{H}_{0i+k,j+k}=\boldsymbol{H}_{0i,j},
\]
then if considered channel is infinite the same property is also satisfied
by Green's function $\boldsymbol{G}_{0}=\left(E_{\mathrm{F}}\boldsymbol{1}-\boldsymbol{H}_{0}\right)^{-1}$
\begin{equation}
\boldsymbol{G}_{0i+k,j+k}=\boldsymbol{G}_{0i,j},\label{eq:transG}
\end{equation}
which shows that knowledge about $\boldsymbol{G}_{0i,j\pm k}$ is
enough to reproduce any blocks of the Green's function. The infinite
device can be truncated with self-energies as in Eq. (\ref{eq:HYG}),
\[
\boldsymbol{\Psi}_{l,m}=\left(E_{\mathrm{F}}\boldsymbol{1}-\boldsymbol{H}\right)^{-1}\boldsymbol{\Gamma}_{l,m}=\boldsymbol{G}\boldsymbol{\Gamma}_{l,m}.
\]
Let us assume that an electron enters the channel in mode $\ket{\boldsymbol{u}_{m,-}}$,
from the left lead $l\equiv L$ (i.e. an electron is injected by the
source vector $\boldsymbol{\Gamma}_{L,m}$, see Eq. \eqref{eq:SourceVec})
and at site $i=0$. Then the scattering wave function at $i$-th slice
is given by 
\begin{equation}
\ket{\boldsymbol{c}_{i,m}}=\boldsymbol{G}_{i,0}\boldsymbol{\Gamma}_{L,m}=\boldsymbol{G}_{i,0}\boldsymbol{\tau}_{L}\left(\boldsymbol{F}_{L,+}-\boldsymbol{F}_{L,-}\right)\ket{\boldsymbol{u}_{L,m,-}}.\label{eq:cm}
\end{equation}
Since the derivation involves only the left lead, we can drop the
subscript $L$ in all the equations below for the sake of clarity.
On the other hand we can write exact formula for the scattering mode
at slice $i$ (pure propagation in the bulk) 
\begin{equation}
\ket{\boldsymbol{c}_{i,m}}=\lambda_{m,-}^{-i}\ket{\boldsymbol{u}_{m,-}}.\label{eq:cme}
\end{equation}
By putting Eq. (\ref{eq:cm}) and Eq. (\ref{eq:cme}) together and
using the matrix notation we get 
\[
\boldsymbol{U}_{-}\boldsymbol{\Lambda}_{-}^{-i}=\boldsymbol{G}_{i,0}\boldsymbol{\tau}\left(\boldsymbol{F}_{+}-\boldsymbol{F}_{-}\right)\boldsymbol{U}_{-},
\]
from which we obtain 
\begin{eqnarray*}
\boldsymbol{G}_{i,0} & = & \boldsymbol{U}_{-}\boldsymbol{\Lambda}_{-}^{-i}\boldsymbol{U}_{-}^{-1}\left[\boldsymbol{\tau}\left(\boldsymbol{F}_{+}-\boldsymbol{F}_{-}\right)\right]^{-1}\\
 & = & \boldsymbol{F}_{-}^{i}\left[\boldsymbol{\tau}\left(\boldsymbol{F}_{+}-\boldsymbol{F}_{-}\right)\right]^{-1},
\end{eqnarray*}
where we used the Eq. \eqref{eq:BlochF}. By setting $i=0$ we obtain
a general expression for the diagonal blocks of the Green's matrix
\begin{equation}
\boldsymbol{G}_{k,k}=\boldsymbol{G}_{0,0}=\left[\boldsymbol{\tau}\left(\boldsymbol{F}_{+}-\boldsymbol{F}_{-}\right)\right]^{-1},\label{eq:Goo}
\end{equation}
which leads to the recursive formula for right off-diagonal elements
$\boldsymbol{G}_{i+k,k}$ 
\begin{equation}
\boldsymbol{G}_{i+k,k}=\boldsymbol{G}_{i,0}=\boldsymbol{F}_{-}^{i}\boldsymbol{G}_{0,0}=\boldsymbol{F}_{-}\boldsymbol{G}_{i-1,0},\,\mathrm{for\, i\geq1.}\label{eq:Gio}
\end{equation}
For the left off-diagonal blocks we get an analogical expression

\begin{equation}
\boldsymbol{G}_{-i,0}=\boldsymbol{G}_{-i+k,k}=\boldsymbol{F}_{+}^{i}\boldsymbol{G}_{0,0}=\boldsymbol{F}_{+}\boldsymbol{G}_{-i+1,0},\,\mathrm{for\, i\geq1.}\label{eq:Gio2}
\end{equation}
The diagonal block matrix $\boldsymbol{G}_{k,k}$ can be directly
computed from Eq. (\ref{eq:Goo}). However, this can be numerically
unstable since both Bloch matrices $\boldsymbol{F}_{\pm}$ and the
coupling matrix $\boldsymbol{\tau}$ can be in general ill-defined
\cite{Rungger2008}. To overcome this problem one may find $\boldsymbol{G}_{k,k}$
by using one of the RGF methods discussed in previous Section . However
we can use the fact that in general $\boldsymbol{G}_{k,k}$ does not
depend on the length of the device and the smallest possible device
which can be described by $2\times2$ block matrix 
\[
\left[\begin{array}{cc}
\boldsymbol{G}_{0,0} & \boldsymbol{G}_{0,1}\\
\boldsymbol{G}_{1,0} & \boldsymbol{G}_{1,1}
\end{array}\right]=\left[\begin{array}{cc}
E_{\mathrm{F}}\boldsymbol{1}-\left(\boldsymbol{h}+\boldsymbol{\Sigma}_{L}\right) & \boldsymbol{\tau}^{\dagger}\\
\boldsymbol{\tau} & E_{\mathrm{F}}\boldsymbol{1}-\left(\boldsymbol{h}+\boldsymbol{\Sigma}_{R}\right)
\end{array}\right]^{-1},
\]
with $\mathbf{h}=\boldsymbol{H}_{0,1,1}$ being the diagonal slice
of the Hamiltonian. From this we can calculate required diagonal element
$\boldsymbol{G}_{0,0}$ as 
\begin{eqnarray}
\boldsymbol{G}_{0,0} & = & \left(\boldsymbol{A}-\boldsymbol{B}\boldsymbol{D}^{-1}\boldsymbol{C}\right)^{-1},\label{eq:Goo2}\\
\boldsymbol{A} & = & E_{\mathrm{F}}\boldsymbol{1}-\left(\boldsymbol{h}+\boldsymbol{\Sigma}_{L}\right)\nonumber \\
\boldsymbol{B} & = & \boldsymbol{\tau}^{\dagger}\nonumber \\
\boldsymbol{C} & = & \boldsymbol{\tau}\nonumber \\
\boldsymbol{D} & = & E_{\mathrm{F}}\boldsymbol{1}-\left(\boldsymbol{h}+\boldsymbol{\Sigma}_{R}\right),\nonumber 
\end{eqnarray}
which we found to be more stable than the direct calculation from
Eq. (\ref{eq:Goo}).

General expression described in this Section can be used to compute
Green's function of translationally invariant devices like graphene
ribbons, carbon nanotubes, quantum wires, straight channels etc. Translational
symmetry of the problem stated in Eq. (\ref{eq:transG}) allows us
to store just one row of the Green's function, hence the memory usage
scales linearly as $(n+1)L^{2}$, where $L$ is the size of the $\boldsymbol{G}_{0,0}$
matrix and $n$ is the number of off-diagonal elements to be computed.
The standard approach requires $n^{2}L^{2}$ elements to be stored,
therefore much larger systems can be stored with this method. The
speed of the algorithm depends mostly on the time needed for computation
of the self-energies, two $L\times L$ matrix inversions in Eq. (\ref{eq:Goo2}),
and $n$ matrix-matrix multiplications defined by equations (\ref{eq:Gio})
and (\ref{eq:Goo2}).

Finally, in this case one can easily compute $\boldsymbol{\Psi}_{M_{L}}^{P}$
(\ref{eq:YPML}) and $\boldsymbol{\overleftarrow{\Psi}}_{M_{L'}}^{P(T)}$(\ref{eq:YPMLprim})
matrices from pure propagation as in Eq. (\ref{eq:cme}).

\subsection{Modular approach to compute Green's function \label{sub:Modular}}

\begin{figure}[H]
\includegraphics[width=1\columnwidth]{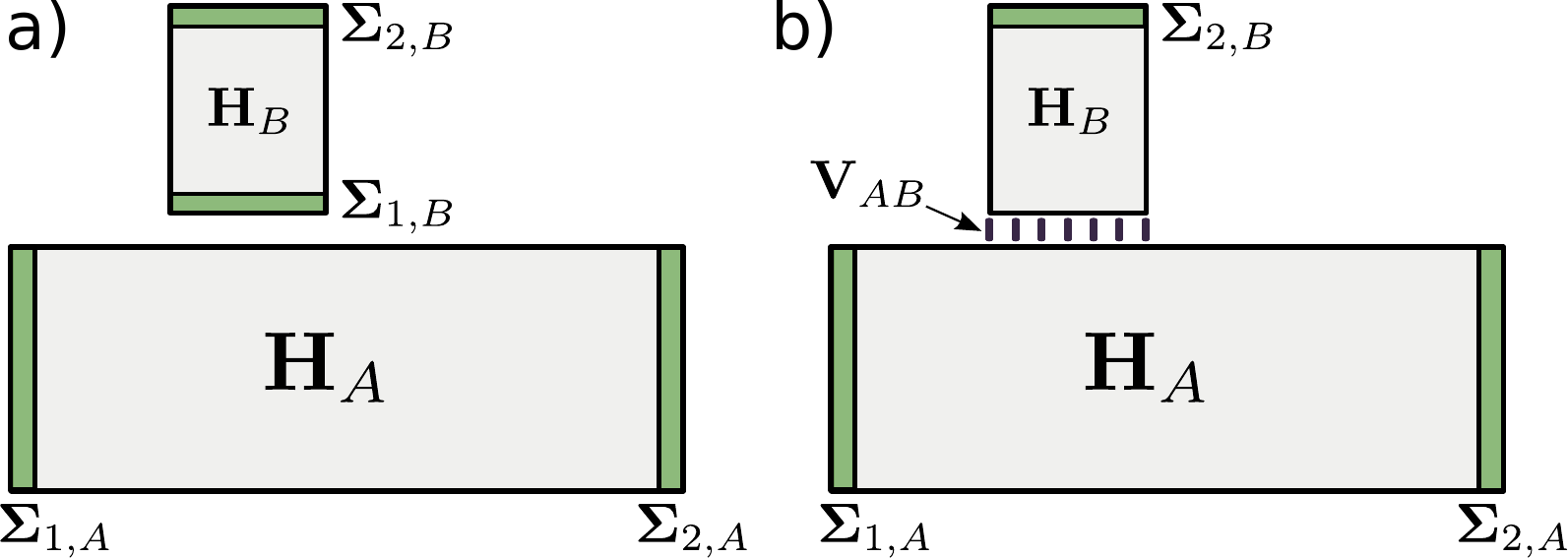}

\caption{\label{fig:modular}a) Schematics  of two uncoupled infinite
channels horizontal A and vertical B. b) System B is glued to system
A with coupling matrix $\mathbf{V}_{AB}$ which removes the self-energy
term from the Hamiltonian B and connects proper sites of both systems. }
\end{figure}

The result derived in the previous Section can be used to construct
efficiently more complicated devices build from translational invariant
blocks connected with proper coupling matrix by using the Dyson equation
similarly as in Ref. \cite{Rotter2003,Sols1989}. As an example we
consider a system created from two channels: a horizontal and a vertical
one (see Fig. (\ref{fig:modular})(a)). Green's function of separated
systems can be computed from Eq. (\ref{eq:Gio}) and (\ref{eq:Gio2}).
The Hamiltonian and the Green's function of uncoupled systems can
be written as 
\[
\mathbf{H}_{0}=\left[\begin{array}{cc}
\boldsymbol{H}_{A} & \mathbf{0}\\
\mathbf{0} & \boldsymbol{H}_{B}
\end{array}\right],\,\mathbf{G}_{0}=\left[\begin{array}{cc}
\boldsymbol{G}_{A} & \mathbf{0}\\
\mathbf{0} & \boldsymbol{G}_{B}
\end{array}\right],
\]
with $\boldsymbol{H}_{X}=E_{F}\mathbf{1}-\left(\boldsymbol{H}{}_{0,X}+\mathbf{\mathbf{\Sigma}}_{1,X}+\mathbf{\mathbf{\Sigma}}_{2,X}\right)$,
with $\boldsymbol{H}{}_{0,X}$ being the Hamiltonian of a closed system.

We can glue both systems with Dyson equation 
\begin{equation}
\mathbf{G}=\mathbf{G}_{0}+\mathbf{G}_{0}\mathbf{V}_{AB}\mathbf{G},\label{eq:Dyson}
\end{equation}
where the coupling matrix $\mathbf{V}_{AB}$ glues selected sites
of system A and B with matrix $\mathbf{\mathbf{\tau}}_{AB}$ and removes
the self-energy term in the lead 1 of systems B (see Fig. (\ref{fig:modular})(b)).
This procedure creates the three terminal device. The $\mathbf{V}_{AB}$
matrix can be mathematically written as

\[
\mathbf{V}_{AB}=\left[\begin{array}{cc}
\boldsymbol{0} & \mathbf{\tau}_{AB}^{\dagger}\\
\mathbf{\mathbf{\tau}}_{AB} & \mathbf{\mathbf{\Sigma}}_{1,B}
\end{array}\right].
\]
Having $\mathbf{V}_{AB}$ and $\mathbf{G}_{0}$ one may compute selected
elements of $\mathbf{G}$ using the standard approach which solves
the Dyson equation (\ref{eq:Dyson}).

\begin{figure}[h]
\includegraphics[width=1\columnwidth]{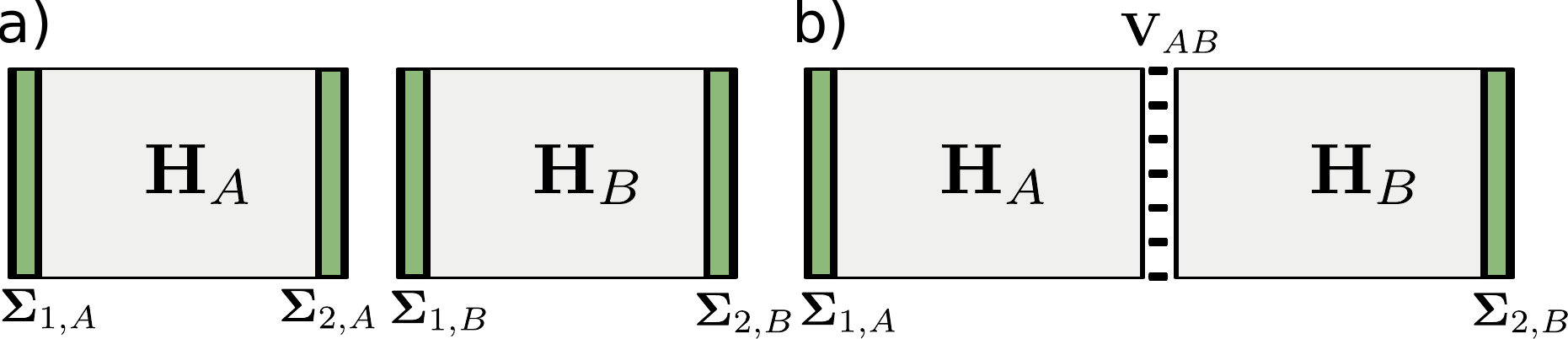}

\caption{\label{fig:modular2}a) Schematic picture of two uncoupled infinite
channels describing different materials. b) After gluing with proper
coupling matrix the systems forms quantum junction. }
\end{figure}

Let us consider another example of two channels created from two different
materials e.g. ferromagnetic A and superconducting B channels or p-n
junction as discussed in the next Section (see Fig. (\ref{fig:modular2})(a)).
Similarly as previously, we can easily compute the Green's function
of separated channels and then glue them together to form a quantum
junction with proper coupling matrix which removes the self-energies
in lead 2 of channel A and lead 1 of channel B (see Fig. (\ref{fig:modular2})(b))

\begin{equation}
\mathbf{V}_{AB}=\left[\begin{array}{cc}
\mathbf{\mathbf{\Sigma}}_{2,A} & \mathbf{\tau}_{AB}^{\dagger}\\
\mathbf{\mathbf{\tau}}_{AB} & \mathbf{\mathbf{\Sigma}}_{1,B}
\end{array}\right].\label{eq:Vab}
\end{equation}

\section{Application to graphene}

The graphene and its transport properties \cite{Neto2009} have been
under an intense investigation for over a decade. The crystal structure
with two non-equivalent triangular sublattices produces a gapless
energy band structure with carriers that behave as massless Dirac
fermions near the charge neutrality point. The presence of two sublattices
and the resulting two non-equivalent Dirac points (K and K' valleys)
forming a symmetric couple under the time inversion leads in particular
to the suppression of the backscattering of chiral carriers by long-range
potentials \cite{Ando2005} and to half plateaux of conductance in
the quantum Hall regime \cite{Novoselov2007,Novoselov2005,Zhang2005,Gusynin2005}.

\subsection{Current equilibration in graphene p-n junction}

In graphene the regions of
of hole or electron conductivity are induced by external
gates, with formation of the n-p junction in the intrinsic material of  homogenous
chemical composition. The n-p junction is
transparent for electrons incident normally \cite{Katsnelson2006}
to the junction (Klein tunneling), and a strong angular dependence
of the transfer probability was used for construction of the Fabry-Pérot interference
in the n-p-n junctions \cite{Shytov2008,Young2009}. However, in the quantum Hall
regime, in high magnetic fields, the n-p junctions serve as waveguides for the
charge currents \cite{Abanin2007,Williams2007}. The current confinement 
at the junction can be classicaly understood as due to the Lorentz force that
act in opposite directions for the carriers of the conduction and valence bands. 
The carriers move along the junction on snake orbits \cite{Carmier2010,Williams2011,Oroszlany2008,Ghosh2008,Zarenia2013}.

\begin{figure}[h]
\includegraphics[width=1\columnwidth]{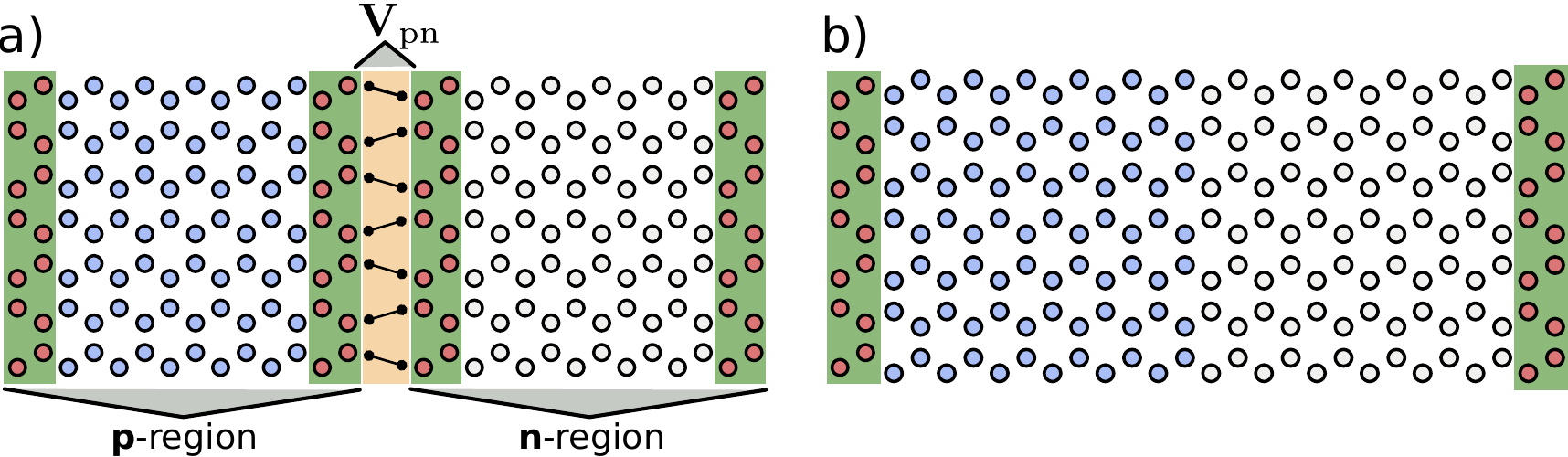}

\caption{\label{fig:e1-pn}a) Sketch of two infinite systems which define $p$
and $n$ parts of the graphene p-n junction. Two systems are then
glued with coupling matrix $\mathbf{V}_{\mathrm{pn}}$. b) The p-n
junction after gluing process. The green areas denote the lead unit
cells. }
\end{figure}

The values of conductance plateaux of the n-p junctions in the quantum
Hall regime can be derived from the assumption of current equilibration,
i.e. mixing of the modes at the contact between the edge and the n-p
junction \cite{Abanin2007}. The mixing is a noncoherent process and
its simulation requires an account taken for dephasing. One of the
procedures \cite{Long2008} uses averaging the conductance through
junction over $N_{\mathrm{samp}}$ different configurations of random
on-site potential introduced on p-n interface i.e. atoms which belong
to the green areas in Fig.\ref{fig:e1-pn}(a). We set the nearest
neighbor carbon-carbon hoping energy to 2.7eV. The potential energy
in p-region is tuned by external gate $\mathbf{V}_{\mathrm{LG}}$.
The number of carbon atoms in the lead cell (see green areas in Figs.
\ref{fig:e1-pn}) is set to 426 which gives ribbon of width $\sim450$\AA{}~for
zigzag edge. The total number of atoms in whole structure is 17466
with total length $\sim50$\AA{}. Note that for this case the size
of the $\mathbf{G}_{\mathrm{int}}$ is $2\times426=852$. The magnetic
field is set to $67$T at which the quantum Hall effect appears. %In order to work with more realistic values of magnetic field one
%should increase the width of the ribbon or simply apply the scaling
%procedure from Ref. \cite{Liu2015Graphene}. 
The on-site energy is uniformly distributed in range $[-W/2,W/2]$
with the disorder strength $W=10$eV. Additionally, for each configuration
we choose randomly 105 of all interface atoms to be affected by on-site
energy.

Green's function of the graphene p-n junction can be constructed by
gluing two infinite systems together. The schematics of the gluing
process is depicted in Fig. \ref{fig:e1-pn}(a). Numerically, the
Green's function of each separated system is calculated with efficient
recursive formulas (\ref{eq:Gio}-\ref{eq:Gio2}) and the coupling
matrix $\mathbf{V}_{\mathrm{pn}}$ defined in Eq. (\ref{eq:Vab})
removes the self-energy matrices and adds hoppings between carbon
atoms at the p-n interface (see black segments in Fig. \ref{fig:e1-pn}(a)).
Then the Green's function matrix $\mathbf{G}_{\mathrm{int}}$ of atoms
at the p-n interface can be computed directly from Dyson equation

\[
\mathbf{G}_{\mathrm{int}}=\left(\mathbf{1}-\mathbf{G}_{\mathrm{pn}}^{0}\mathbf{V}_{\mathrm{pn}}\right)^{-1}\mathbf{G}_{\mathrm{pn}}^{0},
\]
where 
\begin{eqnarray*}
\mathbf{G}_{\mathrm{int}} & = & \left(\begin{array}{cc}
\mathbf{G}_{\mathrm{pp}} & \mathbf{G}_{\mathrm{pn}}\\
\mathbf{G}_{\mathrm{np}} & \mathbf{G}_{\mathrm{nn}}
\end{array}\right),\\
\mathbf{G}_{\mathrm{pn}}^{0} & = & \left(\begin{array}{cc}
\mathbf{G}_{\mathrm{p}} & \mathbf{0}\\
\mathbf{0} & \mathbf{G}_{\mathrm{n}}
\end{array}\right).
\end{eqnarray*}

\begin{figure}[h]
\includegraphics[width=1\columnwidth]{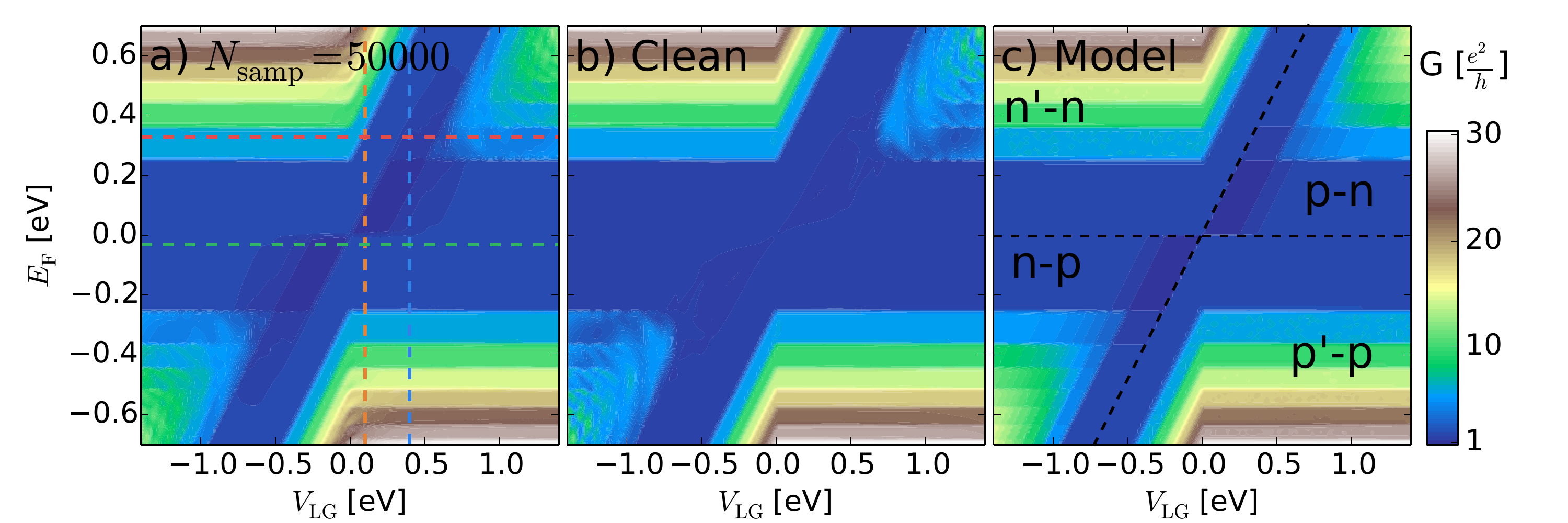}

\caption{\label{fig:e1-pn2}a) Averaged conductance as a function of Fermi
level energy and potential energy $\mathbf{V}_{\mathrm{LG}}$ in the
p-region obtained for $N_{\mathrm{samp}}=10000$ random configuration
of on-site disorder at p-n junction interface. b) Same as (a) but
for clean junction (no averaging). c) Analytical prediction adapted
from Ref. \cite{Levitov2007}.}
\end{figure}

For this configuration the method described above allows for a speedup
by a factor of $\sim$50 in comparison to the standard WFM method
\cite{Sorensen2009,Zwierzycki2008}. The averaged conductance for
$N_{\mathrm{samp}}=10000$ and a clean p-n junction is depicted in
Fig. \ref{fig:e1-pn2}(a) and (b). This can be compared with Fig.
\ref{fig:e1-pn2}(c) obtained from analytical model for fully equilibrated
currents \cite{Levitov2007}. The letters in Fig. \ref{fig:e1-pn2}(c)
denote the different working regimes.

\begin{figure}[h]
\includegraphics[width=1\columnwidth]{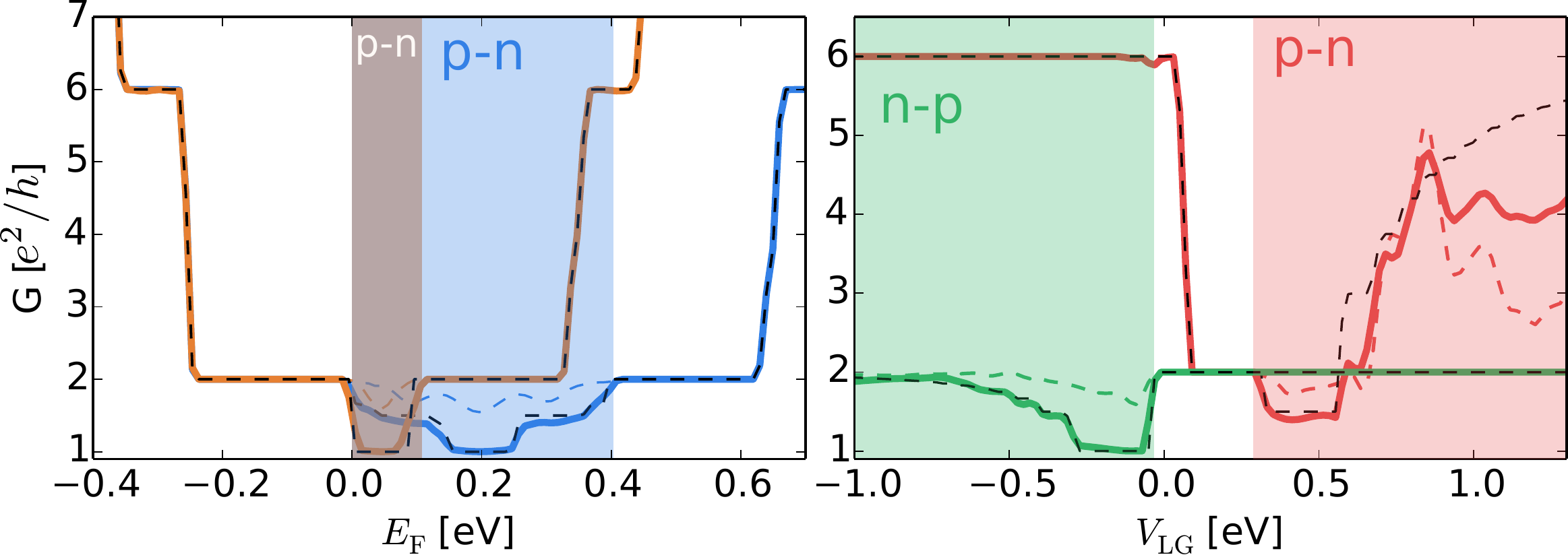}

\caption{\label{fig:e1-cs}The cross sections along vertical a) and horizontal
b) lines in Fig. \ref{fig:e1-pn2}(a). The black dashed lines on each
plot show the analytical prediction for fully equilibrated currents
(Fig. \ref{fig:e1-pn2}(c)). Thick color lines correspond to averaged
conductance from Fig. \ref{fig:e1-pn2}(a). Dashed color lines correspond
to the conductance of clean p-n junction from Fig. \ref{fig:e1-pn2}(b).}
\end{figure}

For unipolar regions (n'-n and p'-p in Fig. \ref{fig:e1-pn2}) we
obtain the same conductance values with or without averaging which
agree with the analytical value of $G=\frac{e^{2}}{h}\min(\nu_{1},\nu_{2})$
\cite{Levitov2007}, where $\nu_{1}$ and $\nu_{2}$ are the Landau
level filling factors ($\nu_{1},\nu_{2}=\pm2,\pm6,\pm10)$ for the
two parts of the gated ribbon. In these conditions the conductance
equal the maximal number of conducting modes for the edge transport
which resist backscattering. For the parameters corresponding to the
n-p junction the conductance plateau are given by \cite{Levitov2007}
$G=\frac{e^{2}}{h}\frac{|\nu_{1}||\nu_{2}|}{|\nu_{1}|+|\nu_{2}|}=1,3/2,3,\dots$.
For the adopted parameters of the random potential only the first
two lowest values are resolved as plateaux (see \ref{fig:e1-cs}).
The applied method \cite{Long2008} requires optimization of the random
disorder parameters for each subsequent conductance plateaux. Note,
that also in the experiment the conductance plateaux at the n-p charge
configuration are less precisely defined than in the unipolar regime
-- see left hand side Fig. 3(c,d) of Ref. \cite{Williams2007}.

\subsection{Magnetic focusing in graphene}

The mean free path of the carriers in graphene reaches several microns
at low temperatures. When an external magnetic field is applied perpendicular
to the graphene plane the electrons move on cyclotron orbits that
can be resolved with the scanning gate microscopy technique (SGM)
\cite{Taychatanapat2013,Morikawa2015,Bhandari2016}. In the SGM measurements
the conductance maps are gathered as functions of the position of
the atomic force microscope tip that acts as a floating gate. The
numerical method described above is a high-performance tool for evaluation
of the conductance maps since (i) the potential of the tip is short
range due to screening the potential of the floating gate by the electron
gas, and (ii) for evaluation of the conductance map one needs to solve
the quantum scattering problem for each location of the tip.

\begin{figure}[h]
\includegraphics[width=1\columnwidth]{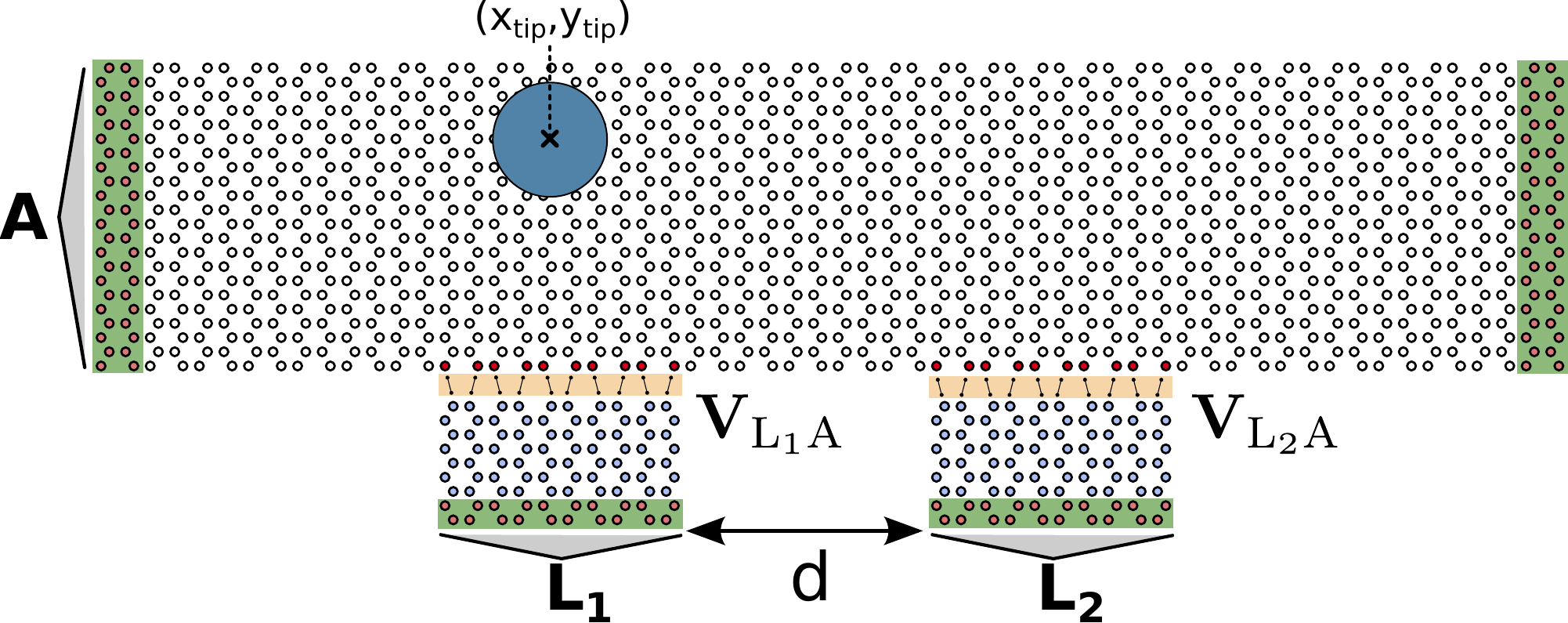}

\caption{\label{fig:e2_mf}The sketch of the magnetic focusing device. The
large reservoir A is build from 191000 atoms and is connected with
two smaller vertical leads ${L}_{1/2}$ with coupling matrices $\mathbf{V}_{m{L}_{1/2},\mathrm{A}}$
separated by distance $d=320$\AA{}. The blue disk shows the SGM tip
influence radius. Only the atoms below the blue area are affected
by the SGM potential.}
\end{figure}

The considered device is built from a large electron reservoir A (see
Fig. \ref{fig:e2_mf}) connected to two smaller leads, the source
${L}_{\mathrm{1}}$ and the drain ${L}_{\mathrm{2}}$ lead (similarly
as in the example in Section \ref{sub:Modular}). The Green's function
of system A is calculated from recursive formulas (\ref{eq:Gio}-\ref{eq:Gio2})
and the Green's functions at the interface of attached leads $L_{\mathrm{1}}$
and $L_{\mathrm{2}}$ are computed with the knitting method \cite{Kazymyrenko2008}.
We define the unperturbed system as a the one constructed from three
areas: A and the leads ${L}_{1/2}$ but without mutual coupling between
them. Then the Green's function is constructed from the Dyson equation.
However, in that case the disorder matrix $\mathbf{V}_{\mathrm{PP}}$
results from a) the SGM tip potential which is modeled as a smooth
disk of radius $d_{\mathrm{tip}}=13$\AA{}~with expression 
\[
V_{\mathrm{tip}}\left(\mathbf{r}\right)=U_{\mathrm{tip}}e^{-\left(\frac{\left|\mathbf{r}-\mathbf{r}_{\mathrm{tip}}\right|}{d_{\mathrm{tip}}}\right)^{8}},
\]
where the center of the tip is located at $\mathbf{r}_{\mathrm{tip}}$
and $U_{\mathrm{tip}}=4$eV and b) the coupling between leads ${L}_{1/2}$
interface sites and A atoms (see black segments in Fig. \ref{fig:e2_mf}).
For each position of the tip we list all the atoms for which condition
$V_{\mathrm{tip}}\left(\mathbf{r}_{\mathrm{atom}}\right)>10^{-3}$eV
is satisfied, resulting in about $300$ atoms on average. The coupling
between ${L}_{1/2}$ and A is introduced at 120 atoms for each both
leads. Hence, the full coupling matrix $\mathbf{V}_{\mathrm{PP}}$
and the reduced Green's function matrices are of size about $420\times420$.
In our simulations the A system is build from 191000 carbon atoms
(with 764 atoms in lead unit cell) and the small leads ${L}_{1/2}$
contains 2970 atoms (with 60 atoms in lead cell) separately. The distance
$d$ between the vertical leads is set to 320 \AA{}. The width and
the length of the A ribbon are $470$\AA{}~and $1060$\AA{}, respectively.
The Fermi energy $E_{\mathrm{F}}$ is set to $0.5$eV. For this set
of parameters we obtain the speedup $\sim15$ in comparison to the
standard method.

\begin{figure}[h]
\includegraphics[width=1\columnwidth]{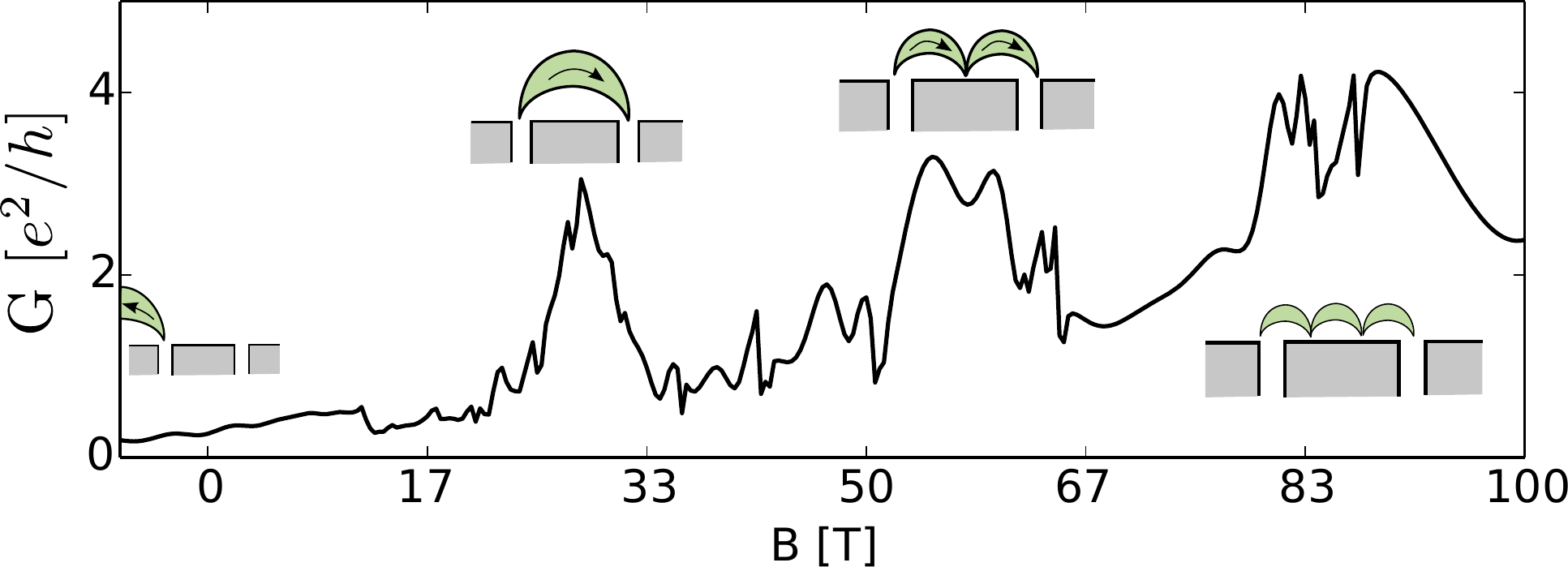}

\caption{\label{fig:e2_s} Conductance between vertical leads as a function
of magnetic field amplitude. The insets indicate the trajectories behind  the
conductance peaks. }
\end{figure}

In Fig. \ref{fig:e2_s} we show the conductance between the source
and drain leads as a function of magnetic field amplitude. The conductance
is calculated with a standard method and the higher conductance means
a higher probability that electron with get from lead ${L}_{1}$ to
${L}_{2}$. Three peaks are clearly visible for $B=27,\,54$, $81$
T.

For $E_{F}=0.5$ eV and $V_{F}=10^{6}$ m/s, the electron density
is $n=\frac{E_{F}^{2}}{\pi(hV_{F})^{2}}=18.31\times10^{1}2$/cm$^{2}$,
and the dynamical electron mass \cite{Bhandari2016} equals $m^{*}=\hbar\sqrt{\pi n}/{V_{F}}=0.087m_{0}$.
Then, the cyclotron diameter is equal to $d_{c}=2m^{*}V_{F}/eB=\frac{1000\mathrm{nmT}}{B}$.
For the values of $B$ corresponding to the first conductance peak
the cyclotron radius is equal to 37.5 nm, which agrees well with the
distance between the axis of the vertical leads ${L}_{1}$ and ${L}_{2}$
that equals 38 nm. We conclude that the peaks correspond to integer
multiples of cyclotron diameters. For magnetic fields $B<0$ the current
is deflected towards the left and hence the current quickly drops
to zero.

\begin{figure}[h]
\includegraphics[width=1\columnwidth]{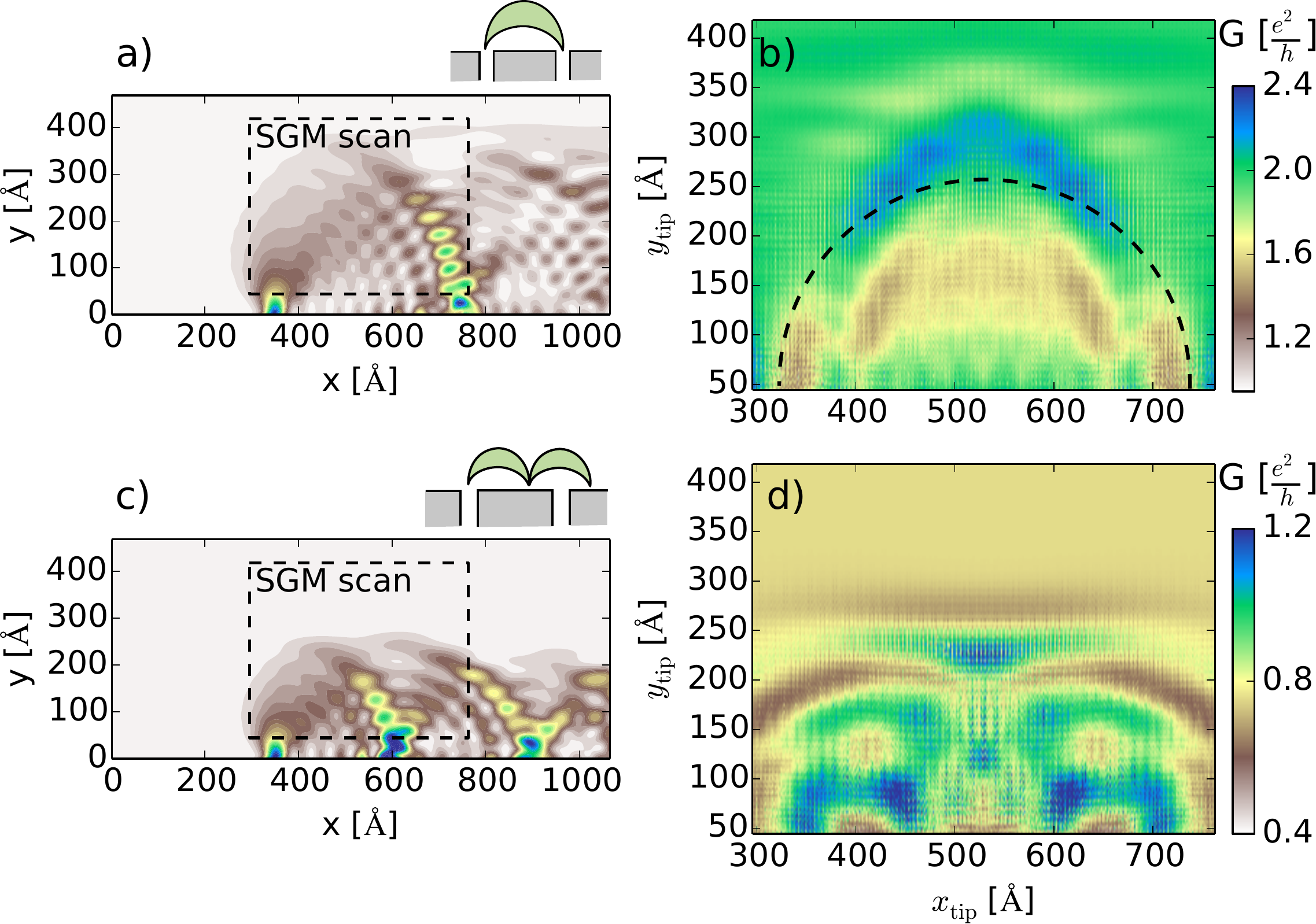}

\caption{\label{fig:e2_sgm}a) The scattering electron density for electron
incoming from lead $\mathbf{L}_{1}$ at $B=27$T. The black dashed
line shows the SGM scan area. b) Simulated SGM image for the case
from (a). Dashed arc corresponds to the classical cyclotron orbit.
c-d) same as (a-b) but obtained for $B=40$T.}
\end{figure}

In Fig. \ref{fig:e2_sgm}(a) we show the scattering electron density
obtained for $B=27.5$T (i.e. the first peak of the conductance in
Fig. \ref{fig:e2_s}). The skipping orbits are clearly seen in the
density plot. In Fig. \ref{fig:e2_sgm}(b) we show the SGM image calculated
with the technique described above which can be compared with the
experimental results given in Fig. 3(b) of Ref. \cite{Bhandari2016}.
The theoretical result of Fig. \ref{fig:e2_sgm}(a) reproduces the
reduced value of conductance when the tip is above the cyclotron orbit
and thus prevents the electrons from passing from $L_{1}$ to $L_{2}$.
Moreover, the present result reproduces the region of increased conductance
when the tip is above the cyclotron orbit, and when it scatters the
electrons to $L_{2}$, and next a ring of reduced conductance, exactly
as observed in Fig. 3(b) of Ref. \cite{Bhandari2016}. To our knowledge
the present result is the first simulation of the magnetic focusing
experiment of Ref. \cite{Bhandari2016} that is based on the solution
of the quantum scattering problem (Ref. \cite{Bhandari2016} used
a classical picture for the interpretation). For completeness in in
Fig. \ref{fig:e2_sgm}(c-d) we show the electron density at a minimum
of the conductance ($B=40$T) with the simulated SGM image. Formation
of a skipping orbit is observed, but with the size that does not coincide
with the distance between the feeding and drain contacts.

\section{Conclusions}

To summarize, we have shown that by projecting the time reversed scattering
wave function onto Lippmann-Schwinger equation we may significantly
reduce the number of Green's function elements needed for computation
of the scattering matrix of arbitrary TB systems in the ballistic
transport approximation.

We have studied the weak perturbation regime for which we have shown
that the first correction to the conductance does not depend on the
Green's function of unperturbed system which resembles the existing
formulas obtained from analytical models. In the case of the one dimensional
wire with delta-like impurity we have shown that the diagonal element
of the Green's function (i.e. local density of states) at the perturbation
site can be extracted from conductance versus impurity strength characteristic.
Additionally, we have discussed the possible applications of our method for
a) current equilibration at the graphene p-n junction interface, obtaining
fractional conductance steps similar to those found in the experiment
\cite{Williams2007}, b) simulation of imaging of the cyclotron orbits
in magnetic focusing experiment with good agreement with Ref. \cite{Bhandari2016}.
For both cases we have obtained significant speedup in comparison
to the standard wave function matching method.

\subsection*{Acknowledgments }

The first author  was supported by National Science Centre according to decision
DEC- 2015/17/N/ST3/02266, by the scholarship of Krakow Smoluchowski Scientific
Consortium from the funding for National Leading Reserch Centre  by
Ministry of Science and Higher Education (Poland) and by the Etiuda
stipend of the National Science Centre (NCN) according to decision
DEC-2015/16/T/ST3/00310. The authors wish to acknowledge Xavier Waintal
and Christoph Groth for valuable and stimulating discussion.
The calculations were supported by PL-Grid Infrastructure.

\section{Appendix}

\subsection{Note on stable calculation of scattering amplitudes.}

After solution of the scattering problem one has to calculate scattering
amplitudes from Eq. (\ref{eq:tm}) and Eq. (\ref{eq:rm}) which involves
inversion of possibly ill-conditioned $\boldsymbol{U}_{l,-}$ matrix
\cite{Rungger2008}. However, one may note that we do not need to
calculate the whole $\mathbf{t}_{l,m}^{l'}$ and $\mathbf{r}_{l,m}^{l'}$
vectors, but only those elements which correspond to the propagating
modes i.e. $\mathbf{t}_{l,m}^{l'}=\left(t_{l,m}^{l',1},\ldots,t_{l,m}^{l',M},t_{l,m}^{l',M+1},\ldots,t_{l,m}^{l',N_{L'}}\right)$,
with $M$ being the number of propagating modes in the lead $l'$.
In order to calculate the first $M$ elements of vector $\mathbf{t}_{l,m}^{l'}$
(the procedure for $\mathbf{r}_{l,m}^{l'}$ is the same) we apply
QL factorization of $\boldsymbol{U}_{l',-}$ matrix 
\begin{equation}
\boldsymbol{U}_{l',-}=\boldsymbol{Q}\boldsymbol{L},\label{eq:Ul}
\end{equation}
with $\boldsymbol{Q}$ being an unitary matrix, and $\boldsymbol{L}$
a lower triangular matrix. The QL factorization can be done even if
$\boldsymbol{U}_{l',-}$ is non-invertible. Now, we can use the fact
that $\mathbf{Q}$ can be easily inverted ($\mathbf{Q}^{-1}=\mathbf{Q}^{\dagger}$)
and $\mathbf{L}$ is a triangular matrix to solve Eq.\eqref{eq:tm}.
However, there is no reason for the first $M\times M$ top-left block
of the $\boldsymbol{L}$ matrix to be well-conditioned and the algorithm
may lead to numerical errors. To avoid this problem we have found
that performing QL factorization of transformed matrix 
\begin{equation}
\boldsymbol{U}_{\mathrm{SVD}}^{\dagger}\boldsymbol{U}_{l',-}=\boldsymbol{Q}'\boldsymbol{L}',\label{eq:Usvd}
\end{equation}
instead of (\ref{eq:Ul}), leads to well ordered triangular matrix
$\boldsymbol{L}'$, where $\boldsymbol{U}_{l',-}=\boldsymbol{U}_{\mathrm{SVD}}\mathbf{S}_{\mathrm{SVD}}\boldsymbol{V}_{\mathrm{SVD}}^{\dagger}$
is the definition of SVD \cite{Rungger2008} and $\mathbf{S}_{\mathrm{SVD}}$
is a diagonal matrix, whose diagonal elements $S_{\mathrm{SVD},k}$
are the singular values which are positive, real and ordered in the
descending order. From Eq. (\ref{eq:Usvd}) we have $\boldsymbol{U}_{l',-}=\boldsymbol{U}_{\mathrm{SVD}}\boldsymbol{Q}'\boldsymbol{L}'$
which we put to the Eq. (\ref{eq:tm}) to get 
\[
\boldsymbol{L}'\mathbf{t}_{l,m}^{l'}=\boldsymbol{Q}'^{\dagger}\boldsymbol{U}_{\mathrm{SVD}}^{\dagger}\boldsymbol{\Psi}_{l,m}^{L'}\equiv\boldsymbol{d}_{m}.
\]
Since $\boldsymbol{L}'$ is lower triangular matrix we may easily
calculate first $M$-th elements with simple recursion without explicit
inversion of full $\boldsymbol{L'}$ matrix 
\[
t_{l,m}^{l',k}=\left(d_{m,k}-\sum_{i=1}^{k-1}L_{k,i}^{'}t_{l,m}^{l',i}\right)/L'_{k,k}.
\]
We find this approach to be more accurate in comparison to direct
inversion of $\boldsymbol{U}_{l',-}$ which in general can be non-invertible.
The reason of the improved stability of Eq. (\ref{eq:Usvd}) comes
from the property of the SVD which order the singular values of the
$\boldsymbol{U}_{l',-}$ matrix in the descending order, hence the
first $M$ rows and columns of $\boldsymbol{U}_{\mathrm{SVD}}^{\dagger}\boldsymbol{U}_{l',-}=S_{\mathrm{SVD}}\boldsymbol{V}_{\mathrm{SVD}}^{\dagger}$
matrix contain contribution of non singular values leading to a more
stable algorithm.

 \bibliographystyle{apsrev4-1-nourl}
\bibliography{referencje}

\end{document}